\documentclass[conference]{IEEEtran}
\IEEEoverridecommandlockouts
\usepackage{cite}
\usepackage{amsmath,amssymb,amsfonts}
\usepackage{graphicx}
\usepackage{textcomp}
\def\BibTeX{{\rm B\kern-.05em{\sc i\kern-.025em b}\kern-.08em
    T\kern-.1667em\lower.7ex\hbox{E}\kern-.125emX}}

\usepackage{amsmath,amsfonts}
\usepackage[ruled,vlined,linesnumbered]{algorithm2e}
\usepackage[noend]{algpseudocode}
\usepackage{array}
\usepackage[caption=false,font=normalsize,labelfont=sf,textfont=sf]{subfig}
\usepackage{textcomp}
\usepackage{stfloats}
\usepackage{url}
\usepackage{verbatim}

\usepackage{todonotes}
\usepackage{enumitem}
\usepackage{tablefootnote}
\usepackage{xcolor}
\usepackage{multirow}
\usepackage{multicol}
\usepackage{soul}
\usepackage{footnote}
\usepackage{colortbl}
\usepackage{tikz}
\usepackage{booktabs}
\usetikzlibrary{arrows,shapes,automata,petri,positioning,calc}

\tikzset{
    place/.style={
        circle,
        thick,
        draw=black,
        fill=gray!30,
        minimum size=6mm,
    },
        state/.style={
        circle,
        thick,
        draw=blue!75,
        fill=blue!20,
        minimum size=6mm,
    },
}

\newcommand*\emptycirc[1][1ex]{\tikz\draw[thin] (0,0) circle (#1);} 
\newcommand*\halfcirc[1][1ex]{%
  \begin{tikzpicture}
  \draw[fill] (0,0)-- (90:#1) arc (90:270:#1) -- cycle ;
  \draw[thin] (0,0) circle (#1);
  \end{tikzpicture}}
\newcommand*\fullcirc[1][1ex]{\tikz\filldraw (0,0) circle (#1);} 

\usepackage{caption}
\usepackage{orcidlink}

\newcolumntype{P}[1]{>{\centering\arraybackslash}p{#1}}
\newcolumntype{M}[1]{>{\centering\arraybackslash}m{#1}}

\newcommand{\ourtool}{\textsf{Valkyrie}}

\begin{document}

\title{\ourtool: A Response Framework to Augment Runtime Detection of Time-Progressive Attacks}

\author{\IEEEauthorblockN{Nikhilesh Singh${^\dag}$\thanks{${^\dag}$The author was associated with the Indian Institute of Technology Madras during the work.}\thanks{This is the authors' version of the paper \ourtool: A Response Framework to Augment Runtime Detection of Time-Progressive Attacks, to appear in the 55th Annual IEEE/IFIP International Conference on Dependable Systems and Networks (DSN) 2025.}\orcidlink{0000-0002-2697-1855}}
\IEEEauthorblockA{\textit{System Security Lab}\\\textit{Technische Universität Darmstadt} \\
Darmstadt, Germany\\ nikhilesh.singh@tu-darmstadt.de
}
\and
\IEEEauthorblockN{Chester Rebeiro\orcidlink{0000-0001-8063-0026}}
\IEEEauthorblockA{\textit{Dept. of Computer Science \& Engg.}\\\textit{Indian Institute of Technology Madras} \\
Chennai, India \\ chester@cse.iitm.ac.in
}
}

\maketitle

\begin{abstract}
A popular approach to detect cyberattacks is to monitor systems in real-time to identify malicious activities as they occur. While these solutions aim to detect threats early, minimizing damage, they suffer from a significant challenge due to the presence of false positives.
 False positives have a detrimental impact on computer systems, which can lead to interruptions of legitimate operations and reduced productivity. Most contemporary works tend to use advanced Machine Learning and AI solutions to address this challenge. Unfortunately, false positives can, at best, be reduced but not eliminated.

In this paper, we propose an alternate approach that focuses on reducing the impact of false positives rather than eliminating them.  We introduce \ourtool{}, a framework that can enhance any existing runtime detector with a post-detection response. \ourtool{} is designed for time-progressive attacks, such as micro-architectural attacks, rowhammer, ransomware, and cryptominers, that achieve their objectives incrementally using system resources. As soon as an attack is detected, \ourtool{} limits the allocated computing resources, throttling the attack, until the detector's confidence is sufficiently high to warrant a more decisive action. For a false positive, limiting the system resources only results in a small increase in execution time. On average, the slowdown incurred due to false positives is less than 1\% for single-threaded programs and 6.7\% for multi-threaded programs. On the other hand, attacks like rowhammer are prevented, while the potency of micro-architectural attacks, ransomware, and cryptominers is greatly reduced. 
\end{abstract}

\begin{IEEEkeywords}
Post-Detection Response, False Positives, Time-Progressive Attacks, Runtime  Attack Detection.
\end{IEEEkeywords}

% \vspace{-0.2cm}
\section{Introduction}\label{sec:intro}

The last few years have seen an unprecedented rise in malware that threatens the security of computing systems. Side-by-side, new malware classes such as ransomware, micro-architectural attacks, cryptominers, and rowhammer have evolved to be extremely potent attack vectors. Ransomware encrypts the contents of a victim's file system and is growing at an alarming rate of 27\% per year~\cite{thales_threat_report2024}. Similarly, with the increased usage of cryptocurrencies, cryptomining malware has also grown significantly, by more than six times in 2023~\cite{sonicwall_threat_report2024}. Micro-architectural attacks~\cite{Yuval:2018:survey} and rowhammer~\cite{Mutlu:2014:Rowhammer}
are new attack vectors that utilize inherent weaknesses in the hardware, making them difficult to defend against. While micro-architectural attacks leak sensitive information through shared hardware resources, rowhammer attacks are capable of flipping bits in memory without accessing them.

A common feature among all of these attacks is that they progress incrementally with execution time. For example, if ransomware executes for a longer duration, it can encrypt more files. Thus, the progress of the ransomware attack increases with execution time. Similarly, if executed for a longer duration, cryptominers are likely to compute more hashes, micro-architectural attacks can leak more information, while rowhammer is likely to flip more bits in memory.
These attacks, which we call {\em time-progressive attacks}, 
achieve their objectives gradually during their execution.

An extensively studied approach to counter time-progressive attacks at runtime is based on the program execution behavior~\cite{Alam:2017:HPCtoRescue,Briongos:2018:cacheShield,Chiapetta:2016:hpcdetection,Gulmezoglu:2019:FortuneTeller,Kulah:2019:SpyDetector,Mushtaq:2018:NightsWatch,Mushtaq:2020:whisper,Mushtaq:2021:transit_guard,Nomani:2015:schedHPC,Payer:2016:HexPADS,Wang:2020:Scarf,Zhang:2016:cloudRadar}.
A typical runtime detector profiles features of executing programs to train a Machine Learning (ML) model~\cite{Ahmed:2021:ransomware_peeler,Alam:2017:HPCtoRescue,Chiapetta:2016:hpcdetection,Gulmezoglu:2019:FortuneTeller,Kulah:2019:SpyDetector,Mushtaq:2018:NightsWatch,Mushtaq:2020:whisper,Mushtaq:2021:transit_guard,Nomani:2015:schedHPC,Wang:2020:Scarf} to identify patterns unique to time-progressive attacks. For example, hardware performance counters (HPCs) present in processors~\cite{Alam:2017:HPCtoRescue,Demme:2013:feasibility,Gulmezoglu:2019:FortuneTeller,Kulah:2019:SpyDetector,Mushtaq:2018:NightsWatch,Mushtaq:2020:whisper,Mushtaq:2021:transit_guard,Nomani:2015:schedHPC,Wang:2020:Scarf}, system-level APIs~\cite{Ahmed:2021:ransomware_peeler,Karapoola:2024:Sundew,MSprocmontool}, and network activity~\cite{Karapoola:2024:Sundew,wiresharktool} are periodically measured and used by trained models to detect an ongoing attack.
Such detection approaches can be easily adapted to support new attacks, while at the same time, applied at various levels of the computing stack, such as the network, Operating System (OS), and  hardware~\cite{Karapoola:2024:Sundew}.

A limitation of these ML-based detectors is that they are susceptible to false positives. As a result, benign programs are classified malicious. Existing works address these false positives by using more sophisticated detection models~\cite{Chiapetta:2016:hpcdetection,Gulmezoglu:2019:FortuneTeller,Karapoola:2024:Sundew}. For example, Gulmezoglu {\em et al.}~\cite{Gulmezoglu:2019:FortuneTeller} use unsupervised deep learning techniques, while Karapoola {\em et al.}~\cite{Karapoola:2024:Sundew} employ a multi-level mixture of experts model for detection. 
However, none of these solutions can completely eliminate false positives in real-world deployments~\cite{Das:2019:sokHPC,Kosasih:2024:HPCs_FP,Zhou:2018:HPCmythorfact}. For instance,~\cite{Alam:2017:HPCtoRescue} implements perceptrons and has up to 7\% false positives, while~\cite{Mushtaq:2021:transit_guard} which uses more sophisticated machine learning models, has under 3\% false positives for detecting micro-architectural attacks.

False positives in attack detection can adversely impact operations. For instance, a high number of false positives in the alerts raised by the detector at the retail network Target, led to alerts being ignored by the security team. This eventually resulted in one of the most prominent cyber attacks in recent years~\cite{AlAhmadi:2022:TargetFP}. Instead of alerting users about ongoing attacks, another option is to terminate programs that are classified malicious. This would result in a large number of falsely classified programs being prematurely terminated. In some cases~\cite{Nomani:2015:schedHPC,Zhang:2016:cloudRadar}, the detected programs can be migrated to a different execution environment, resulting in significant overheads. Thus, there is a pressing need to design efficient responses, following a detection.

Post-detection response strategies for time-progressive attacks have two critical requirements. {\sf R1:} The response should throttle attacks and minimize their progress. {\sf R2:} The response should minimally affect the execution of benign programs when falsely classified.  Throttling attacks and minimizing their progress requires timely detection and termination. In contrast, reducing the adverse effects of benign programs necessitates improving the efficacy of detection, primarily
minimizing false positives. However, the higher detection efficacy would require more complex detection models or a larger number of runtime measurements, causing delays in thwarting attacks. 

Most existing works, unfortunately, target satisfying {\sf R1} but not {\sf R2}, hampering the usability of the solution. For instance, even a low false-positive rate, of say, 3\%~\cite{Mushtaq:2021:transit_guard}, would mean that 3 out of every 100 programs, on average, would be terminated prematurely.  Addressing this limitation requires either the elimination of false positives or a reduction in the impact of false positives. Since the elimination of false positives with present-day ML models is challenging,  
it is essential to focus on minimizing the impact of false positives. Thus, we pose the following research question in this work: {\em can we design post-detection response mechanisms for runtime detectors that satisfy} {\sf R1} {\em and} {\sf R2}{\em, thwarting attacks while minimizing the adverse impacts of false positives?}

 A characteristic feature of all time-progressive attacks is that they require system resources, such as CPU, memory, network, or the file system, for successful execution.  Limiting these resources would result in slowing down the attack's progress.  For example, limiting the time a ransomware executes in the CPU,  restricting the memory allocated to the process, or restricting file system access can slow down the rate at which files are encrypted. We use this observation to present~\ourtool{}, a framework to augment detectors with post-detection response strategies \footnote{The relevant code and data associated with~\ourtool{} are maintained at \url{https://bitbucket.org/casl/valkyrie/src/main/}}. Designed to handle the adverse impacts of false positives,~\ourtool{} augments a detector that periodically provides an inference. Programs that are detected malicious are terminated only if the detection has sufficient \textit{detection efficacy}. If the detection efficacy is not achieved, detected programs are slowed down by throttling system resources needed for their execution. For an attack, the slowdown would reduce its progress, while for a false positive, there is a temporary slowdown. This is less adverse than termination. Thus, the choice of detection efficacy affects the slowdown of false positives as well as the rate at which attacks are stymied. We define detection efficacy as a metric that quantifies the capability of the detector to classify benign and malicious program behavior. It can be represented by model evaluation metrics such as F1-score and false positive rate (FPR). Users can specify constraints regarding the detection efficacy, based on the application. Once this efficacy is reached, the program is terminated only if it continues to be classified as malicious. On the other hand, if the program is classified as benign -- a false positive -- then its resources are gradually restored. Thus,~\ourtool{} provides a response mechanism ensuring that benign programs do not face premature termination ({\sf R2}) while attacks are throttled ({\sf R1}).

Following are the key contributions of the paper.

\begin{itemize}
    \item We present~\ourtool{}, a response framework that augments detection countermeasures for time-progressive attacks to minimize the impacts of false positives while ensuring that attacks get throttled. \ourtool{} slows down the progress of attacks by restricting system resources and terminates a detected process only when the detector achieves sufficient detection efficacy specified by the user.
    \item  The implementation of~\ourtool{} requires minimal changes and can be plugged into any detector. We demonstrate~\ourtool{} by augmenting multiple detectors against various time-progressive attacks. We present case studies involving various micro-architectural attacks, such as Prime+Probe on the L1 Data~\cite{Osvik:2006:cacheFlushing}, L1 Instruction~\cite{Aciimez:2010:instncache}, and LLC caches~\cite{Maurice:2017:cjag,Yuval:2018:Mastik};  L1 and TLB Evict+Time attacks~\cite{Gras:2018:TLB,Osvik:2006:cacheFlushing}; and Fill-and-Forward Timed Speculative Attack on Load-Store Buffer~\cite{Chakraborty:2022:TSA}. Similarly, we also evaluate~\ourtool{} with rowhammer~\cite{Mutlu:2014:Rowhammer,rowhammer:2020:github}, ransomware~\cite{bware_ransomware:2024,gonnacry_ransomware:2024,open_ransomware:2024,raasnet_ransomware:2024,randomware_ransomware:2024}, and cryptominers~\cite{Papadopoulos:2018:cryptominers}.
    \item We mathematically quantify the slowdowns induced by~\ourtool{} when benign processes are falsely classified. We present an empirical evaluation of these slowdowns with multiple benign benchmark suites, including SPEC-2017~\cite{spec2017}, SPEC-2006~\cite{SPEC2006}, SPECViewperf-13~\cite{specviewperf13}, STREAM~\cite{McCalpin:2007:stream}, and multi-threaded SPEC-2017~\cite{spec2017}. The average slowdowns are 1\% for single-threaded and 6.7\% for multi-threaded programs across different evaluation systems. Additionally,~\ourtool{} also supports a configurable upper limit for maximum slowdowns due to resource throttling.  
\end{itemize}

The rest of the paper is organized as follows. Section~\ref{sec:background} presents the necessary background on time-progressive attacks and their detection. We provide an overview of existing post-detection approaches in Section~\ref{sec:related_work}. Section~\ref{sec:motivation_and_challenges} describes the motivation for~\ourtool{} and presents 
an overview of the framework. In Sections~\ref{sec:valkyrie_design} and~\ref{sec:results}, we discuss the design of~\ourtool{} in detail and demonstrate the utility of~\ourtool{} by augmenting several detectors for various time-progressive attacks as case studies, respectively. Section~\ref{sec:caveats} 
presents a discussion on the limitations of~\ourtool{}, while Section~\ref{sec:conclusions} concludes the paper.

\begin{table}[!t]
\centering
\footnotesize
    \caption{\small Existing runtime detection countermeasures and their post-detection responses along with the reported false positives. [Req: Rquirement; \emptycirc: requirement not satisfied, \halfcirc: requirement partially satisfied, \fullcirc: requirement satisfied]. }
    %\vspace{-0.1cm}
    \begin{tabular}{|M{0.2\columnwidth}
|M{0.3\columnwidth}|M{0.02\columnwidth}|M{0.02\columnwidth}|M{0.2\columnwidth}|}
    \hline
  \multirow{2}{*}{\shortstack{\bf Post-Detection\\{\bf Response}}} & \multirow{2}{*}{\textbf{Papers}} & \multicolumn{2}{c|}{\textbf{Req.}} & \multirow{2}{*}{\shortstack{\bf False positives\\{\bf reported}}} \\
   \cline{3-4}
    & & {\sf R1} & {\sf R2} & \\
    \hline   
   \multirow{14}{*}{Not Specified}  &  Alam {\em et al.}~\cite{Alam:2017:HPCtoRescue} & \multirow{14}{*}{\emptycirc} & \multirow{14}{*}{\emptycirc}
 & 5-7\% \\
\cline{2-2}\cline{5-5}
 & Briongos {\em et al.}~\cite{Briongos:2018:cacheShield} &  &    & 1.6-4.3\% \\
 \cline{2-2}\cline{5-5}
 & Chiapetta {\em et al.}~\cite{Chiapetta:2016:hpcdetection} &  &   & Not reported \\
 \cline{2-2}\cline{5-5}
 & Gulmezoglu {\em et al.}~\cite{Gulmezoglu:2019:FortuneTeller} &  &  & 0.21\% \\
  \cline{2-2}\cline{5-5}
 & Mushtaq {\em et al.}~\cite{Mushtaq:2018:NightsWatch} &  &  & 1-30\% \\
\cline{2-2}\cline{5-5}
 & Mushtaq {\em et al.}~\cite{Mushtaq:2020:whisper} &  &  & 5\%\\
 \cline{2-2}\cline{5-5}
 & Wang {\em et al.}~\cite{Wang:2020:Scarf} &  &  & upto 13.6\% \\
 \cline{2-2}\cline{5-5}
& Karapoola {\em et al.}~\cite{Karapoola:2024:Sundew} &  &  & 0.01\% \\
\cline{2-2}\cline{5-5}
& Ahmed {\em et al.}~\cite{Ahmed:2021:ransomware_peeler} &  &  & 0.58\% \\
\cline{2-2}\cline{5-5}
& Vig {\em et al.}~\cite{vig:2018:rowhammer_detection} &  &  & 1\% \\
\cline{2-2}\cline{5-5}
& Pott {\em et al.}~\cite{Pott:2023:Cryptominer_detection} &  &  & 0.2\% \\
\cline{2-2}\cline{5-5}
& Tahir {\em et al.}~\cite{Tahir:2017:cryptominer} &  &  & 0.25\% \\
\cline{2-2}\cline{5-5}
& Mani {\em et al.}~\cite{Mani:2020:DL_cryptominer} &  &  & 0.2-3.8\% \\

    \hline 
   Warning & Kulah {\em et al.}~\cite{Kulah:2019:SpyDetector} & \emptycirc & \emptycirc & Not reported\\
   \hline 
   
\multirow{2}{*}{Migration} &  Zhang {\em et al.}\cite{Zhang:2016:cloudRadar}& \multirow{2}{*}{\fullcirc} & \multirow{2}{*}{\halfcirc} & Not reported \\
\cline{2-2}\cline{5-5}
& Nomani {\em et al.}~\cite{Nomani:2015:schedHPC} &  &  & Not reported\\
   \hline 
   \multirow{2}{*}{Termination} & Mushtaq {\em et al.}~\cite{Mushtaq:2021:transit_guard}  & \fullcirc  & \halfcirc  & 1-3\%\\
   \cline{2-5}
   & Payer~\cite{Payer:2016:HexPADS} & \halfcirc & \halfcirc  & Not reported \\
   \hline
   \multirow{2}{*}{\shortstack{DRAM\\responses}} & Aweke {\em et al.}~\cite{Aweke:2016:anvil} & \multirow{2}{*}{\fullcirc}  & \multirow{2}{*}{\fullcirc}  & 1\% \\
\cline{2-2}\cline{5-5}
& Yaglik{\c{c}}i {\em et al.}~\cite{Yaglici:2021:BlockHammer} &  &  & 0.01\%\\
   \hline 
   Systematic throttling and eventual termination & \ourtool{} (this paper; augmented with any detector)  & \fullcirc  & \fullcirc  & Same as augmented detector \\
   \hline 
   \end{tabular}
   \label{tab:related_work}
\end{table}

\section{Background}\label{sec:background}

\subsection{Time-Progressive Attacks}\label{sec:time_progressive_attacks}

Time-progressive attacks use system resources such as CPU, memory, network, kernel APIs, or the filesystem to achieve their objectives incrementally with execution. For example, consider  cryptominers~\cite{Papadopoulos:2018:cryptominers}. These attacks aim to mine cryptocurrency, which involves a computationally intensive challenge, such as recovering the input of a hash function from a given output. The longer the cryptominer executes, the more challenges can be solved to mine cryptocurrency.  Similarly, micro-architectural attacks rely on the execution time
measurements to glean information. When given
 a longer time to execute, these attacks can collect more measurements and, in turn, leak more bits of information. 
Restricting the system resources such as CPU time, memory, network and filesystem can, thus, impede the progress of time-progressive attacks. 
In this paper, we make use of this dependence of the attack's progress on the system resources to design~\ourtool{}. 

\subsection{Runtime Detection of Attacks}\label{sec:attack_detection}

Runtime detection of time-progressive attacks typically involves detection models trained on measurements representing program behavior. These measurements are captured using tools such as the Linux Perf tool~\cite{linuxperftool:2019}, which captures the CPU's hardware performance counters (HPCs); Microsoft ProcMon~\cite{MSprocmontool}, which captures the operating system (OS) events; or Wireshark~\cite{wiresharktool}, which captures network behavior. Machine Learning (ML) based detection models such as SVM~\cite{Mushtaq:2020:whisper}, XGBoost~\cite{Karapoola:2024:Sundew} and artificial neural networks (ANNs)~\cite{Nomani:2015:schedHPC} are trained on such measurements from benign and malicious programs. After training, the detector is deployed to classify processes using the measurements that are continuously captured for each process at runtime.

Several existing works have demonstrated the efficacy of such runtime detectors to identify time-progressive attacks~\cite{Alam:2017:HPCtoRescue,Briongos:2018:cacheShield,Chiapetta:2016:hpcdetection,Gulmezoglu:2019:FortuneTeller,Kulah:2019:SpyDetector,Mushtaq:2018:NightsWatch,Mushtaq:2020:whisper,Mushtaq:2021:transit_guard,Nomani:2015:schedHPC,Payer:2016:HexPADS,Wang:2020:Scarf,Zhang:2016:cloudRadar}. Unfortunately, all of these approaches have been shown to be susceptible to false positives~\cite{Das:2019:sokHPC,Kosasih:2024:HPCs_FP, Zhou:2018:HPCmythorfact}, which impact their usability. For this work, the comparative assessment of different types of measurements and detectors is not relevant. Rather, we are interested in post-detection response strategies. We discuss this in the next section.

\section{Existing Post-Detection Approaches}\label{sec:related_work}

Table~\ref{tab:related_work} presents the post-detection response strategies followed in existing runtime detection countermeasures along with the reported false positive rate (FPR) of each detector. The FPR represents the benign programs that are affected by the response. We evaluate these post-detection responses with respect to the requirements {\sf R1} and {\sf R2} described in Section~\ref{sec:intro}. Most existing runtime detectors~\cite{Ahmed:2021:ransomware_peeler,Alam:2017:HPCtoRescue,Briongos:2018:cacheShield,Chiapetta:2016:hpcdetection,Gulmezoglu:2019:FortuneTeller,Karapoola:2024:Sundew,Mani:2020:DL_cryptominer,Mushtaq:2018:NightsWatch,Mushtaq:2020:whisper,Pott:2023:Cryptominer_detection,Tahir:2017:cryptominer,vig:2018:rowhammer_detection,Wang:2020:Scarf} for time-progressive attacks do not specify any post-detection response. Works such as Kulah {\em et al.}~\cite{Kulah:2019:SpyDetector} respond with a warning to the user once a process is classified malicious. Since handling the detection requires a vigilant user, it is challenging to guarantee either of the requirements {\sf R1} or {\sf R2} with these approaches. 

A promising response strategy for micro-architectural attacks is to migrate the detected process to a different CPU core or another virtual machine (VM). While this approach satisfies both {\sf R1} and {\sf R2}, it has two key limitations. First, migration is applicable only against contention-based micro-architectural attacks~\cite{Nomani:2015:schedHPC,Zhang:2016:cloudRadar} such as cache attacks in cloud infrastructure~\cite{percival:05,Irazoqui:2015:cacheAttackVM}. 
Second, migration of processes across VMs can induce high overheads. For instance, a process that is falsely classified malicious throughout its execution by~\cite{Nomani:2015:schedHPC} can incur up to 50\% performance overheads due to migration across CPU cores.

The detection countermeasure presented in 
 Payer~\cite{Payer:2016:HexPADS} requires users to select between termination or a reduction in the execution priority of attacks like rowhammer~\cite{Mutlu:2014:Rowhammer}. While termination satisfies {\sf R1} but not {\sf R2}, reducing the execution priority may not satisfy {\sf R1} as it can allow attacks to execute endlessly. Further, relying on users to respond after detection also affects the consistency of the approach. Mushtaq {\em et al.}~\cite{Mushtaq:2021:transit_guard} discuss the challenge of false positives in micro-architectural attack detection and attempt to address it with a post-detection response. Rather than terminating a process right after detection, this solution terminates a process only when it is classified malicious three times consecutively. While such an approach reduces the number of benign processes that are falsely terminated from 5\% to under 3\% ({\sf R2}), the choice of waiting for three consecutive classifications is arbitrary and can not be generalized across detectors.

 The authors in~\cite{Aweke:2016:anvil} and~\cite{Yaglici:2021:BlockHammer} present responses to refresh the DRAM after detecting a rowhammer attack. While these approaches satisfy both {\sf R1} and {\sf R2}, the response specifically targets rowhammer and is not applicable to other attacks. Unlike the post-detection responses in existing works,~\ourtool{} can augment any runtime detector targeting a wide range of time-progressive attacks, making it scalable and generalizable. Once the solution is deployed,~\ourtool{} does not require any involvement of the user. Further, the responses from~\ourtool{} are not arbitrary, but are designed based on the threat level of a process and the specifications given by the user for detection efficacy and performance slowdowns.

\section{Motivation and Overview}\label{sec:motivation_and_challenges}

The design of~\ourtool{} is based on two key observations. The first is that the efficacy of detection models increases over time with the number of captured runtime measurements of process behavior~\cite{Kulcher:2021:sample_length}. Second, time-progressive attacks are impacted by the availability of system resources, as discussed in Section~\ref{sec:time_progressive_attacks}. 
In this section, we analyze these two observations to motivate the design of~\ourtool{}. We then describe the 
high-level overview of the design.

\begin{figure*}[!t]
\captionsetup[subfloat]{farskip=0.5pt,captionskip=0.5pt}
\centering
\begin{tabular}{cc}

\subfloat[\small F1-Score.]{\includegraphics[width=0.4\textwidth]{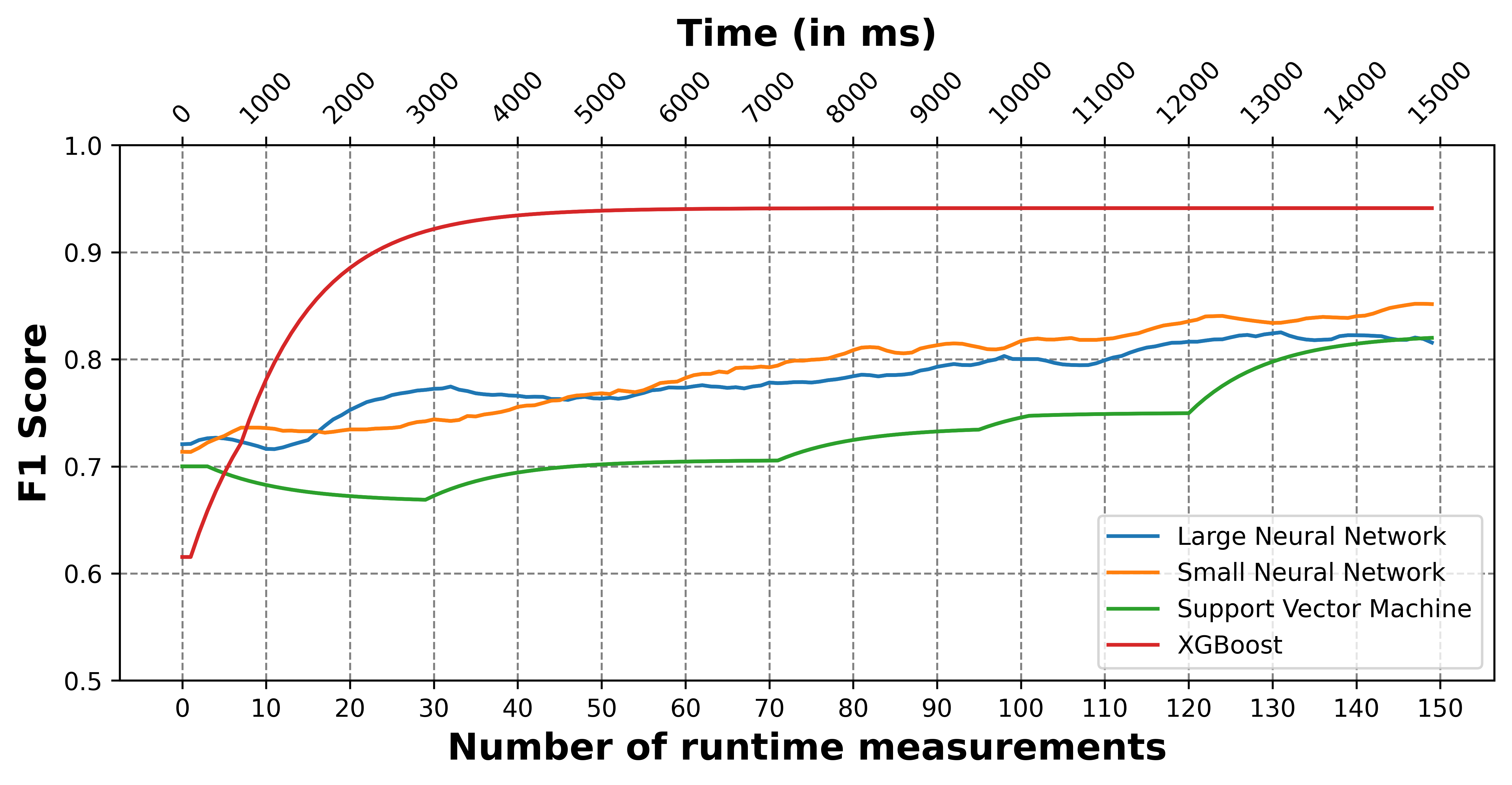}\label{fig:f1_vs_time}} &
\subfloat[\small False Positive Rate.]{\includegraphics[width=0.4\textwidth]{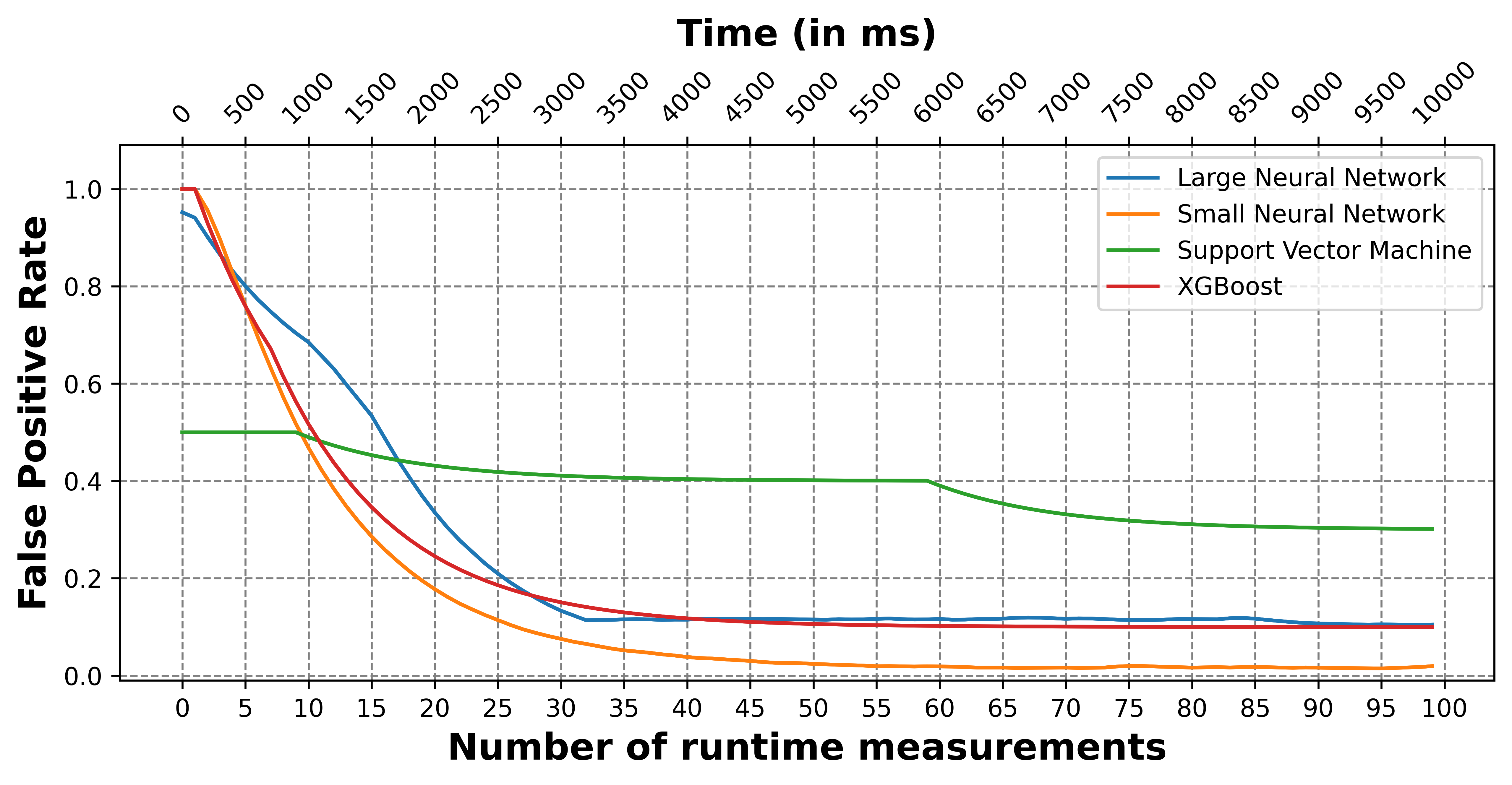}\label{fig:fpr_vs_time}} 
\\
\end{tabular}

% \vspace{-0.2cm}
\caption{\small Detection efficacy of various models similar to existing works with respect to the number of runtime measurements captured by the detector and elapsed time. The user of a detection solution can specify the desired level of detection efficacy.}\label{fig:detection_efficacy_vs_time}
\end{figure*}

\subsection{Detection Efficacy Over Time}\label{sec:efficacy_vs_measurements}
 To detect time-progressive attacks at runtime, measurements representing program behavior are captured periodically. A detector augmented with~\ourtool{} uses these measurements to provide an inference for each process at periodic intervals called a \textit{measurement epoch} or an \textit{epoch}. Thus, with every passing epoch, the detector accumulates a larger number of measurements to classify the process behavior. For most detection models, the detection efficacy improves with the measurements captured. Fig.~\ref{fig:detection_efficacy_vs_time} presents the detection efficacy of popular detection models with respect to the number of measurements captured. We use metrics such as F1-Score (Fig.~\ref{fig:f1_vs_time}) and the False Positive Rate (FPR) (Fig.~\ref{fig:fpr_vs_time}) to represent the detection efficacy. The analysis includes small and large artificial neural networks (ANNs) similar to~\cite{Alam:2017:HPCtoRescue,Chiapetta:2016:hpcdetection,Gulmezoglu:2019:FortuneTeller,Mushtaq:2021:transit_guard}, a Support Vector Machine (SVM) model as used in~\cite{Mushtaq:2018:NightsWatch,Karapoola:2024:Sundew}, and an XGBoost ensemble as deployed in~\cite{Karapoola:2024:Sundew}. The small ANN has one hidden layer of 4 nodes, while the large ANN has two hidden layers of 8 nodes each. 
 All the evaluated detectors in Fig.~\ref{fig:detection_efficacy_vs_time} use HPC measurements to detect ransomware programs.  In these experiments, we use 67 ransomware programs from various open-source repositories~\cite{bware_ransomware:2024,gonnacry_ransomware:2024,open_ransomware:2024,raasnet_ransomware:2024,randomware_ransomware:2024}. Each model provides an inference after every measurement  ({\em i.e.,} every epoch has one additional measurement). The ANNs take a time series of HPC measurements as input to classify programs. On the other hand, the SVM and XGBoost models classify each measurement individually and infer program behavior based on the classification of majority of these measurements. For the purpose of our discussion, the comparison across these detectors is not relevant. Rather, we are interested in the change in detection efficacy over time. We observe that the efficacy of each detector improves with the number of measurements. For instance, the F1-Score of the small ANN is 0.7 with 5 measurements, which increases to 0.8 with 75 measurements. For a typical HPC monitoring tool~\cite{linuxperftool:2019} that captures hardware events every 100ms, 
 these measurements would require 7 seconds.

We can utilize the trend that detection efficacy improves with time (Fig.~\ref{fig:detection_efficacy_vs_time}) to determine the number and duration of measurements required to achieve a specified level of efficacy. For instance, to get an F1-Score of more than 0.9, the XGBoost detector would need 23 measurements, thus, 2.3 seconds. Similarly, an FPR of less than 10\% on the same model would require at least 5 seconds of measurements. While accumulating measurements for a longer time improves the detection efficacy, it also enables the attacks to progress further. For example, by the time the ANN improves its F1-Score to 0.8, ransomware encrypting data at a rate of 11.67MB/s~\cite{bware_ransomware:2024}, can corrupt 81MB of data. Similarly, a micro-architectural attack that gleans data at 40KB/s~\cite{Maurice:2017:cjag}, can extract 300KB. Thus, we need a solution that enables the detector to accumulate measurements for a longer time, thereby attaining sufficient detection efficacy while, at the same time, thwarting the progress of the attacks.

\begin{figure*}[!t]
    \centering
    \includegraphics[width=0.9\textwidth]{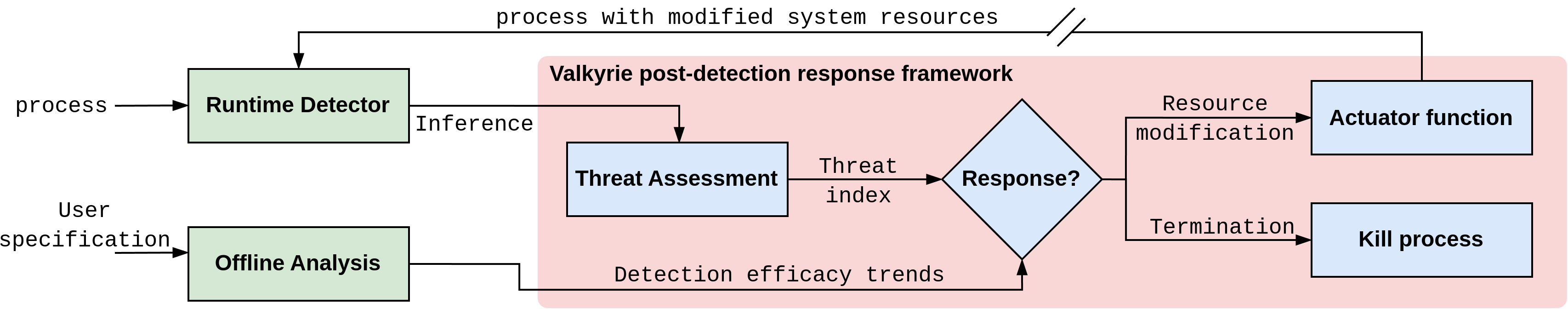}
    \caption{\small An overview of~\ourtool{} post-detection response framework augmenting a runtime detector. In an offline phase, users can specify the detection efficacy required based on the application, and~\ourtool{} determines the number of measurements required to achieve it. The detector provides an inference for a process periodically during execution, which is used by~\ourtool{} to determine the threat index of the process. Based on the threat index and the number of measurements,~\ourtool{} responds with a modification in available resources for the process or termination.}
    \label{fig:valkyrie_cycle}
\end{figure*}

\subsection{Resource Availability and Attack Progress}\label{sec:resource_vs_progress}

% {\color{red} One way to the stuff in previous section is to evaluate..}

To design a post-detection response framework that thwarts the attack progress while the detector attains a specified efficacy, we evaluate the dependence of time-progressive attacks on system resources. Consider an attack that \textbf{(a)} recursively opens files present in the system, \textbf{(b)} computes the hash of each file, and \textbf{(c)} sends the hash and the contents of each file over the network to a colluding server. The progress of this attack can be measured by the number of bytes transmitted to the server. This is a time-progressive attack (Section~\ref{sec:background}), as the number of hashes computed and bytes transmitted increases with time. The attack uses resources such as the disk to access files, memory for data and instructions, network to transmit the file contents, and CPU for execution. 
Table~\ref{tab:resources_vs_attack_progress} demonstrates the effects of variation in available resources on the progress of the program. 
We control the system resources available to the program using management features in the Linux kernel~\cite{cgroup:2024}.

% \vspace{-0.1cm}
{\flushleft \bf CPU time.} We quantify the CPU usage based on the fraction of time for which a program executes on the CPU core. We control this CPU time using a Cgroup-based utility~\cite{cgroup:2024}. As shown in Table~\ref{tab:resources_vs_attack_progress}, reducing the CPU share to 1\% can slow down the rate of progress by 99.7\% with respect to the default. 

% \vspace{-0.1cm}
{\flushleft \bf Memory.} We measure the available memory given to the program relative to the maximum memory space used by the program without any restrictions. We modify the available memory for a program by creating a Cgroup and assigning limits on the usable memory. This ensures that the program never exceeds the specified memory usage by forcibly invalidating the associated pages.  We observe that a reduction in the available memory significantly brings down the attack progress, as shown in Table~\ref{tab:resources_vs_attack_progress}. For instance, the attack's rate of progress slows down by 99.99\% when we reduce the available memory to 93.6\% of the required memory.

\begin{table}[]
    \centering
    \scriptsize
    \caption{\small The rate of progress of the example time-progressive attack that recursively computes the hash of the victim's files and transmits the contents to a colluding server, with variations in the available system resources.}
    \begin{tabular}{M{0.14\columnwidth}|M{0.14\columnwidth}|M{0.14\columnwidth}|M{0.16\columnwidth}|M{0.16\columnwidth}}
    \hline 
     \multirow{4}{*}{\textbf{Resource}}  & \multicolumn{2}{c|}{\textbf{Availability}} &   \multicolumn{2}{c}{\textbf{Rate of Progress}} \\
       \cline{2-5}
       & \textbf{Value} & \textbf{\% of default} & \textbf{Bytes transmitted (KB/second)} & \textbf{\% slowdown} \\
        \hline 
        \multirow{4}{*}{\textbf{CPU}} & 100\% [default] & 100\%  & 225.7 & $-$ \\
        \cline{2-5}
        & 90\% & 90\% & 206.1 & 8.7\% \\
        \cline{2-5}
        & 50\% & 50\% & 123.6 & 45.2\%\\
        \cline{2-5}
        & 1\% & 1\% & 0.6 & 99.7\%\\
        \hline
        \multirow{4}{*}{\textbf{Memory}} & 4.7M [default] & 100\% &  225.7 & $-$ \\
        \cline{2-5}
        & 4.6M & 93.6\% & 0.087 & 99.96\% \\
        \cline{2-5}
        & 4.4M & 89.4\% & 0.013 & 99.99\% \\
        \hline 
        \multirow{4}{*}{\textbf{Network}} & 1024G [default] & 100\%  & 225.7 & $-$ \\
        \cline{2-5}
        & 512G & 50\% & 200 & 11.4\% \\
        \cline{2-5}
        & 512M & $10^{-3}$\% & 56.63 & 74.9\% \\
        \cline{2-5}
        & 512K & $10^{-6}\%$ & 0.05 & 99.98\% \\
        \hline 
        \multirow{4}{*}{\textbf{Filesystem}} & 100 file/s [default] & 100\% & 225.7  & $-$ \\
        \cline{2-5}
        & 90 files/s & 100\% &200.1 & 11.3\% \\
        \cline{2-5}
        & 50 files/s & 50\% & 113.7 & 49.6\% \\
        \cline{2-5}
        & 1 file/s & 1\% & 2.2 & 99\% \\
        \hline 
    \end{tabular}
    \label{tab:resources_vs_attack_progress}
    %\vspace{-0.6cm}
    \end{table}

% \vspace{-0.1cm}
{\flushleft \bf Network.} Limiting the network bandwidth affects the transmission of file contents by the attack program to the server. To regulate the available network bandwidth, we use the Cgroup feature and specify upper bounds for the usage of network bandwidth. As shown in Table~\ref{tab:resources_vs_attack_progress}, 
halving the network bandwidth slows down the rate of progress by 11.4\%.

% \vspace{-0.1cm}
{\flushleft \bf Filesystem.} Modifying the rate of filesystem access affects file reading for hash computations and transmission. We control this rate by keeping track of the files opened and using signals to pause and resume execution. Changes in the rate of file accesses affect the rate of progress proportionally (Table~\ref{tab:resources_vs_attack_progress}).

We observe that different system resources influence the attack progress at varying rates. For instance, the available CPU time and the rate of file access affect the rate of progress in a proportional manner. Availability of the network bandwidth has a linear effect on the progress. On the other hand, limiting the available memory has a non-linear and sharp effect on the attack's progress. Thus, throttling the available memory can restrict the attack progress sharply while we can have a graceful decline in progress by throttling the CPU time and file accesses.

% \vspace{-0.2cm}
\subsection{\ourtool{}: An Overview }\label{sec:challenges}

We have two important takeaways from the previous sections. First, a runtime detector 
needs to accumulate measurements for longer durations in order to improve its detection efficacy (Section~\ref{sec:efficacy_vs_measurements}). Second, restricting system resources can impede time-progressive attacks (Section~\ref{sec:resource_vs_progress}). 
Based on these two observations, we design our post-detection response framework,~\ourtool{}.

Fig.~\ref{fig:valkyrie_cycle} provides the high-level overview of~\ourtool{}. For each executing process, the detector augmented with~\ourtool{}, provides a periodic inference. The response to these inferences depends on the expected level of detection efficacy needed by the target system. For example, critical systems necessitate termination of attacks as early as possible. A higher FPR in such systems is more easily tolerated. From Fig.~\ref{fig:detection_efficacy_vs_time}, this would require that the detector provides its response based on fewer measurements. 
General purpose systems, on the other hand, are more sensitive to false positives. Therefore, detectors should terminate attacks in general purpose systems after observing larger number of measurements. 

In \ourtool{}, users can specify the expected detection efficacy. Based on this input \ourtool{} computes the number measurements needed to achieve the specified efficacy. While the detector accumulates the required number of measurements,~\ourtool{} proportionally throttles the processes classified malicious by regulating the system resources, thereby reducing the attack's progress as described in Section~\ref{sec:resource_vs_progress}.

To implement~\ourtool{}, we address the following questions.

{\flushleft \bf Q1: Which system resources to throttle?} The pattern of resource utilization varies from one time-progressive attack to another. For instance, ransomware has a significant dependence on the CPU as well as the file system. On the other hand, cryptominers are entirely CPU-dependent and do not use the filesystem. Thus,~\ourtool{} monitors processes at runtime to ensure that the system resources critical to the attack progress are throttled.

{\flushleft \bf Q2: How to throttle the system resources?} Once the resources to throttle are identified, we need a mechanism to (a) quantify and (b) regulate the resources. To quantify the resources available to a process,~\ourtool{} calculates a \textit{threat index} for the process. The threat index is a value between $0$ and $100$ that quantifies the maliciousness of the process. A higher threat index implies a higher confidence in the process being malicious. The calculation of the threat index for a process at any instant is based on 
all the previous inferences from the detector. The threat index increases when the detector classifies the process as malicious and decreases when it is classified as benign. An \textit{actuator function} uses the threat index to regulate the share of system resources available to the process. For example, the actuator function can use Linux kernel features to throttle CPU, memory, network, and the filesystem, as discussed in Section~\ref{sec:resource_vs_progress}.

{\flushleft \bf Q3: How long to throttle and when to terminate?} \ourtool{} regulates the resources available to the process until the detector attains the detection efficacy specified by the user. After this, the process can be terminated if classified as malicious by the detector, or its resources are restored if it is detected as a false positive (classified benign). 

% \vspace{-0.1cm}
\section{The~\ourtool{} response mechanism}\label{sec:valkyrie_design}

In this section, we discuss the design of~\ourtool{} with a formal description. Consider a detector $\mathcal{D}$ augmented with~\ourtool{}, which uses the runtime measurements to classify process behavior. Let ${\tt N}^*$ be the number of measurements required by $\mathcal{D}$ to achieve the detection efficacy specified by the user. ${\tt N}^*$ is determined offline by learning the changes in detection efficacy with respect to the number of measurements, as discussed in  Section~\ref{sec:efficacy_vs_measurements}.
Let ${\tt t}$ be a process executing on the system with ${\tt N}_{i}^{\tt t} (< {\tt N}^*)$ measurements captured till the start of the $i$-th epoch. Based on these measurements, the detector $\mathcal{D}$ classifies ${\tt t}$ as malicious or benign in the $i$-th epoch, represented by $\mathcal{D}({\tt t}, i) = \{\text{malicious}, \text{benign} \}$.

The set $R_i^{\tt t}$ represents the share of system resources available to process ${\tt t}$ in the $i$-th epoch, such that,

% \vspace{-0.4cm}
\begin{align}
R_i^{\tt t}  = \{ r_{\text{CPU}}^{\tt t}(i), r_{\text{mem}}^{\tt t}(i),  r_{\text{nw}}^{\tt t}(i), r_{\text{fs}}^{\tt t}(i)\} \enspace,
\end{align}

 where $r_{\text{CPU}}^{\tt t}(i),r_{\text{mem}}^{\tt t}(i),r_{\text{nw}}^{\tt t}(i)$ and $ r_{\text{fs}}^{\tt t}(i)$ represent the share of CPU time, memory, network, and filesystem, respectively available to the process $\tt t$ in the $i$-th epoch. 
 
 \ourtool{} utilizes the inference $\mathcal{D}({\tt t}, i)$ and the user specification (${\tt N}^*$) to assess the threat posed by process $\tt t$. The threat assessment is used to determine a response for $\mathcal{D}({\tt t}, i)$. For instance, we can increase, decrease, or maintain the share of system resources (a subset of $R_i^{\tt t}$) available to ${\tt t}$, thus affecting the execution of the process. 

 % current textfloatsep
 \newlength{\textfloatsepsave} \setlength{\textfloatsepsave}{\textfloatsep} 
 % set textfloatsep to 0
 \setlength{\textfloatsep}{0pt} 
\begin{algorithm}[!t]
\DontPrintSemicolon
{\bf Global:} 
A process ${\tt t}$; ${\tt state}({\tt t})$: state of process ${\tt t}$; $\mathcal{D}$(${\tt t}$, $i$): online detector's inference in $i$-th epoch; $R_{i}^{\tt t}$: share of resources available to ${\tt t}$ in $i$-th epoch; ${\tt N}_{i}^{\tt t}$: measurements captured for ${\tt t}$ till $i$-th epoch; ${\tt N}^*$:  number of measurements required to satisfy user specification; ${\tt clamp}(x) = \max(0, \min(x, 100)).$ \label{lin:clamp_description}\\
{\bf Initial State:} {\tt t} is executing; ${\tt state}({\tt t})=$normal; $i=0; {\tt P}_{i}^{\tt t} = {\tt C}_{i}^{\tt t} = {\tt T}_{i}^{\tt t} = {\tt N}_i^{\tt t} = 0;$ \\
\Begin{
    \While{${\tt t}$\emph{ is executing}}{
    \While{${\tt N}_i^{\tt t} < {\tt N}^*$}{
        $i = i+1$ and  Update ${\tt N}_i^{\tt t}$\\
        Get the inference $\mathcal{D}({\tt t}, i)$\label{lin:get_inference}\\
        \If{$\mathcal{D}({\tt t}, i) == \emph{malicious}$}{\label{lin:penalty_start}
        ${\tt state}({\tt t})$ = suspicious\\
        ${\tt P}_{i}^{\tt t} = {\tt clamp}(\mathcal{F}_p({\tt P}_{i-1}^{\tt t}))$ \label{lin:clamp1}\\
        ${\tt C}_{i}^{\tt t} = {\tt C}_{i-1}^{\tt t}$ \&
         ${\tt T}_i^{\tt t} = {\tt T}_i^{\tt t} + {\tt P}_i^{\tt t}$}\label{lin:penalty_end}
        }
        \Else{
        \If{${\tt state}({\tt t})$ == \emph{suspicious}}{\label{lin:compensation_start}
        ${\tt C}_{i}^{\tt t} = {\tt clamp}(\mathcal{F}_c({\tt C}_{i-1}^{\tt t}))$\label{lin:clamp2}\\
        ${\tt P}_{i}^{\tt t} = {\tt P}_{i-1}^{\tt t}$ \&
        ${\tt T}_i^{\tt t} = {\tt T}_i^{\tt t} - {\tt C}_i^{\tt t}$\label{lin:compensation_end}
        }
        }
        ${\tt T}_{i}^{\tt t} = {\tt clamp}({\tt T}_{i}^{\tt t})$ \label{lin:clamp3}\\
        \If{${\tt T}_i^{\tt t} == 0$}{${\tt state}({\tt t})$ = normal}
        $\Delta {\tt T}_{i,1}^{\tt t} = {\tt T}_{i}^{\tt t} - {\tt T}_{i-1}^{\tt t}$\label{lin:actuator_start}\\
        $R_{i}^{\tt t} = \mathcal{A}(R_{i-1}^{\tt t}, \Delta {\tt T}_{i,1}^{\tt t})$\label{lin:actuator_end}\\
        
    }
    ${\tt state}({\tt t})$ = terminable\label{lin:terminable_start}\\
    $i = i+1$\\
    \If{$\mathcal{D}({\tt t}, i) == \emph{benign}$}{$\mathcal{A}_{\text{reset}}(R_{i-1}^{t})$\tcp{restore ${\tt t}$} }
    \Else{
    Terminate process ${\tt t}$\label{lin:terminate}
    }
    }

\caption{ Execution of process ${\tt t}$ with a detector $\mathcal{D}$ augmented with the~\ourtool{}.}\label{algo:valkyrie_key_idea}
\end{algorithm}

\subsection{Threat Assessment with~\ourtool{}}\label{sec:threat_assessment}

The threat index for the process ${\tt t}$ in the $i$-th epoch, ${\tt T}_{i}^{\tt t}$, is a quantification of the detector's confidence in the process being malicious. A threat index of $0$ implies that the process is benign and has no restrictions on the system resources, while a threat index of $100$ would result in the maximum restrictions on resources. The value of the threat index in an epoch is based on the history of the detector's inferences till that epoch. To determine the threat index,~\ourtool{} maintains two metrics for each process, namely, penalty (${\tt P}_i^{\tt t}$) and compensation (${\tt C}_i^{\tt t}$). Before the detector accumulates the required number of measurements (${\tt N}^*$), the threat index increases by the penalty metric each time the detector classifies $\tt t$ as malicious (Line~\ref{lin:penalty_end} in Algorithm~\ref{algo:valkyrie_key_idea}). If the behavior of the process $\tt t$ improves and the detector classifies it as benign, the threat index decreases by the compensation metric (Line~\ref{lin:compensation_end} in Algorithm~\ref{algo:valkyrie_key_idea}).

To determine the values of ${\tt P}_i^{\tt t}$ and ${\tt C}_i^{\tt t}$,~\ourtool{} uses two configurable functions.
The penalty assessment function 
$(\mathcal{F}_p({\tt P}_i^{\tt t}))$ takes in the penalty value of the previous epoch and increases the penalty value if the process is classified malicious
(Lines~\ref{lin:penalty_start}-\ref{lin:penalty_end} in Algorithm~\ref{algo:valkyrie_key_idea}). Similarly, the compensation assessment function $(\mathcal{F}_c({\tt C}_i^{\tt t}))$ increases the compensation metric if ${\tt T}_i^{\tt t} > 0$ and the process is classified benign (Lines~\ref{lin:compensation_start}-\ref{lin:compensation_end} in Algorithm~\ref{algo:valkyrie_key_idea}). To restrict the values of ${\tt P}_i^{\tt t}$, ${\tt C}_i^{\tt t}$ and ${\tt T}_i^{\tt t}$ between $0$ and $100$, we use a ${\tt clamp}()$ function (Lines~\ref{lin:clamp_description}, \ref{lin:clamp1}, \ref{lin:clamp2}, and \ref{lin:clamp3} Algorithm~\ref{algo:valkyrie_key_idea}). Both these functions can have several possible realizations, such as incremental $({\tt P}_{i}^{\tt t} = \mathcal{F}_p({\tt P}_{i-1}^{\tt t}) = {\tt P}_{i-1}^{\tt t} + 1 )$, linear $(\mathcal{F}_p({\tt P}_{i-1}^{\tt t}) = a {\tt P}_{i-1}^{\tt t} + b, \text{where } a \text{ and } b \text{ are constants})$, or exponential $(\mathcal{F}_p({\tt P}_{i-1}^{\tt t}) = 2^i {\tt P}_{i-1}^{\tt t} + 1)$. Based on these functions, the penalty and compensation metrics can grow at varying rates, which, in turn, influences the threat index (${\tt T}_i^{\tt t}$) and the throttling of the process.

% reset the textfloatsep
\setlength{\textfloatsep}{\textfloatsepsave}

\begin{figure}[!t]
    \centering
     \resizebox{0.7\columnwidth}{!}{%
\begin{tikzpicture}[node distance=2cm and 1cm,>=stealth',auto, every place/.style={draw}]
    \node [place] (S1) {\textbf{normal}};
    \coordinate[node distance=1.1cm,left of=S1] (left-S1);
    \coordinate[node distance=1.1cm,right of=S1] (right-S1);

    \draw[->, thick] (left-S1) -- (S1);

    \node [place] (S2) [below=of S1] {\textbf{suspicious}};
    
    \node [place] (S3) [node distance=5cm, right of=S1] {\textbf{terminable}};

    \node [state, text=,accepting by double] (S4) [below=of S3] {\textbf{terminated}};

    \path[->] (S1) edge [bend right=15] node [swap, align=center, left, rotate=0] {${\tt T}_i^{\tt t} > 0,$\\ ${\tt N}_i^{\tt t} < {\tt N^*}$} (S2);
    
    \path[->] (S2) edge [bend right=15] node [swap, align=center, right, rotate=0, yshift=0, xshift=0]{${\tt T}_i^{\tt t} = 0,$\\ ${\tt N}_i^{\tt t} < {\tt N^*}$} (S1);
    
    \path[->] (S2) edge [loop below, looseness=5, in=-75, out=-105] node [swap, align=center, left, rotate=0, yshift=10pt, xshift=-10pt] {${\tt T}_i^{\tt t} > 0,$\\ ${\tt N}_i^{\tt t} < {\tt N^*}$} (S2);
    
    \path[->] (S2) edge node [swap, align=center, right, rotate=35, yshift=-8pt, xshift=-12pt]{${\tt N}_i^{\tt t} \geq {\tt N^*}$} (S3); 
    
    \path[->] (S1) edge node [swap, align=center, right, rotate=0, yshift=8pt, xshift=-18pt]{${\tt N}_i^{\tt t} \geq {\tt N^*}$} (S3);
    
     \path[->] (S1) edge [loop above, looseness=5, in=65, out=105] node [swap, align=center, left, xshift=-10pt, yshift=-10pt]{${\tt T}_i^{\tt t} = 0,$\\${\tt N}_i^{\tt t} < {\tt N^*}$} (S1);

     \path[->] (S3) edge node [swap, align=center, right, rotate=90, yshift=1pt, xshift=-30pt]{$\mathcal{D}(t,i)$\\$=\text{malicious}$ \\or\\ ${\tt t}$ completes} (S4);

       \path[->] (S3) edge [loop above, looseness=5, in=77, out=105] node [swap, align=center, left, rotate=0, xshift=-8pt , yshift=-12pt] {${\tt t}$ is executing\\ and $\mathcal{D}({\tt t}, i)=\text{benign}$} (S3);
\end{tikzpicture}
        }
        
    \caption{\small The state transitions of a process ${\tt t}$ with~\ourtool{}. The process starts in the normal state (${\tt T}_i^{\tt t} = 0$) and transitions to a suspicious state if it gets classified as malicious ($\mathcal{D}({\tt t},i) = \text{malicious}$, thus ${\tt T}_i^{\tt t} > 0$). The process ${\tt t}$ can remain in the suspicious state (${\tt T}_i^{\tt t} > 0$) or return to a normal state (${\tt T}_i^{\tt t} = 0$) based on its execution behavior. Once the detector accumulates the number of measurements to achieve the detection efficacy specified by the user (${\tt N}_i^{\tt t} \geq {\tt N^*}$), the process switches to the terminable state from normal or suspicious. The process ${\tt t}$ in terminable state gets terminated if the detector classifies it as malicious ($\mathcal{D}({\tt t},i) = \text{malicious}$) or if ${\tt t}$ completes execution.}
    \label{fig:process_states}
\end{figure}
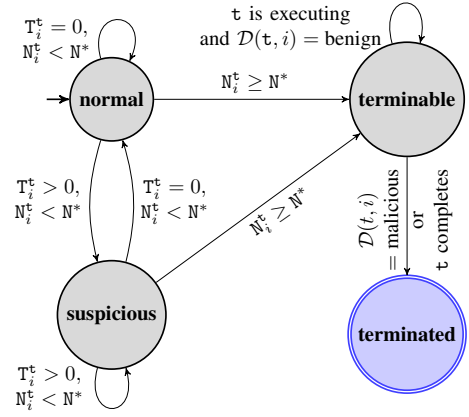

Based on the number of measurements and the threat index, \ourtool{} divides the execution of a process into four possible states, as shown in Fig.~\ref{fig:process_states}. Each process starts in the \textit{normal} state (threat index ${\tt T}_i^{\tt t} = 0$). A process continues to execute in the normal state if the detector does not classify it as malicious in any epoch. Until the user specified detection efficacy is satisfied (${\tt N}_i^{\tt t} < {\tt N}^*$), an increase in the threat index transitions the process to the \textit{suspicious} state. In this state, the threat index of the process ${\tt t}$ in the $i$-th epoch, (${\tt T}_i^{\tt t}$) determines the rate at which the process gets thwarted or recovers. A process can transition from the suspicious to the normal state if the process behavior improves and the threat index falls to zero. We typically observe this in the case of a false positive. Once the user specified detection efficacy is satisfied (${\tt N}_i^{\tt t} \geq {\tt N}^*$), then the process transitions to the \textit{terminable} state. The process transitions to the \textit{terminated} state if the detector classifies it malicious or the execution of the process is complete.

\subsection{Throttling Resources with~\ourtool{}}\label{sec:resource_throttling}

In the suspicious state, the resources available to a process are determined based on the threat index (${\tt T}_i^{\tt t}$). \ourtool{} incorporates an \textit{actuator function} $(\mathcal{A})$ to identify the system resources used by a process and regulate them based on the threat index.  This function $\mathcal{A}(R_{i-1}^{\tt t}, \Delta {\tt T}_{i,1}^{\tt t})$ takes in the share of resources from the previous epoch, $R_{i-1}^{\tt t}$ and the change in threat index in the current epoch $ \Delta {\tt T}_{i,1}^{\tt t}$, to output the updated share of resources available to the process. It ensures a reduction and improvement in the share of available resources with an increase or decrease in the threat index, respectively (Lines~\ref{lin:actuator_start}-\ref{lin:actuator_end} in Algorithm~\ref{algo:valkyrie_key_idea}). The design of $\mathcal{A}$ depends on the resource to throttle. For instance, a possible actuator function to regulate the  CPU time can work by modifying the OS scheduler such that processes with higher threat index are scheduled for shorter durations. Another possibility is to monitor the process execution and use SIGSTOP and SIGCONT signals to pause and resume execution, respectively. Such an actuator function can induce restrictions on the CPU time of a process, similar to utilities like cpulimit~\cite{cpulimit:2024}. Similarly, an actuator can control the available memory, network, and filesystem resources to process using Linux kernel features, as shown in Section~\ref{sec:resource_vs_progress}.

{\flushleft \bf Termination with~\ourtool{}.} 
Once the process ${\tt t}$ transitions to the terminable state and the detector classifies it as benign, the function $\mathcal{A}_{\text{reset}}$ removes all the restrictions on available resources for the process, restoring ${\tt t}$ to its default resources. On the other hand, the process is terminated when the detector classifies it as malicious.

\subsection{Quantifying Slowdowns with~\ourtool{}}\label{sec:slowdowns}

As the progress function of time-progressive attacks depends on the available resources, we define a function $\mathrm{B}_i^{\tt t}(R_i^{\tt t})$ to represent the progress of ${\tt t}$ in the $i$-th epoch. For time-progressive attacks, the precise value of this function depends on the objectives of the attack. For instance, the attack progress can be quantified as the number of bits gleaned by a micro-architectural attack~\cite{Aciimez:2010:instncache,Chakraborty:2022:TSA,Gras:2018:TLB,Maurice:2017:cjag,Osvik:2006:cacheFlushing,Yuval:2018:Mastik}, bits flipped in memory by the rowhammer attack~\cite{Mutlu:2014:Rowhammer}, bytes encrypted by ransomware~\cite{bware_ransomware:2024,gonnacry_ransomware:2024,open_ransomware:2024,raasnet_ransomware:2024,randomware_ransomware:2024,wannacry_blog:2017}, or the hashes computed by a cryptominer~\cite{Papadopoulos:2018:cryptominers}. 

Let the detector $\mathcal{D}$ require $K$ epochs to capture ${\tt N}^*$ measurements for process $\tt t$. If ${\tt t}$ is a time-progressive attack, the progress of ${\tt t}$ in $K$ epochs without~\ourtool{} can be given as,

% \vspace{-0.4cm}
 \begin{equation}
    \text{Attack progress in $\mathrm{K}$ epochs without \ourtool{}} = \sum_{i=0}^{\mathrm{K}-1} \mathrm{B}_i^{\tt t}({R}_i^{\tt t})\label{eqn:attack_progress}
\end{equation}

which, for instance, can be the total number of bits gleaned by a micro-architectural attack. 

Assuming the attack is suspicious state, in $K$ epochs the attack progress with~\ourtool{} is given by, 

% \vspace{-0.6cm}
\begin{align}
    \text{Att} & \text{ack progress in $K$ epochs with \ourtool{}} \nonumber\\
    & = \mathrm{B}_0^{\tt t}({R}_0^{\tt t}) + \sum_{i=1}^{K-1} \mathrm{B}_{i}^{\tt t}\Big(\mathcal{A}({R}_{i-1}^{\tt t}, \Delta {\tt T}_{i,1}^{\tt t}) \Big) \enspace,  \label{eqn:attack_progress_solution}
\end{align}

where $\Delta {\tt T}_{i,1}^{\tt t} =  {\tt T}_i^{\tt t} - {\tt T}_{i-1}^{\tt t}$.
Equations~\ref{eqn:attack_progress} and~\ref{eqn:attack_progress_solution} gives the effective slowdown ($S({\tt t})$) of the process $\tt t$ in percentage due to~\ourtool{}. 

% \vspace{-0.4cm}
\begin{align}
 S({\tt t})= \Bigg(1- \frac{\mathrm{B}_0^{\tt t}({R}_0^{\tt t}) + \sum_{i=1}^{K-1} \mathrm{B}_{i}^{\tt t}(\mathcal{A}\Big({R}_{i-1}^{\tt t}, \Delta {\tt T}_{i,1} \Big)}{\sum_{i=0}^{\mathrm{K}-1} \mathrm{B}_i^{\tt t}({R}_i^{\tt t})}\Bigg) \times 100
\end{align}
% \vspace{-0.3cm}

Thus, the throttling of an attack is dependent on the value of $K$, the threat index, which in turn depends on the penalty and compensation assessment functions $\mathcal{F}_p \text{ and } \mathcal{F}_c$, and the actuator function $\mathcal{A}$. A 0\% slowdown indicates no modification of the available resources by~\ourtool{} and is ideal for benign processes. A slowdown of 100\% implies that the attack progress halts completely. 

Let us understand slowdowns with the example attack described in Section~\ref{sec:resource_vs_progress}. Consider a detector that requires a minimum of $15$ epochs to satisfy the user specified detection efficacy ({\em i.e.,} ${\tt N}^* = 15$) with an incremental penalty and compensation assessment function. Thus, each time the detector classifies the attack as malicious, the penalty increases by 1, and the threat index increases by the penalty value. The actuator in this example drops the CPU share by 10\% for every increase in the threat index (the minimum CPU share is 1\%). If the detector classifies the attack as malicious in every epoch, the attack would incur a slowdown of  
79.6\% before reaching the terminable state (15 epochs). 
A benign process can also incur slowdowns due to false positives. 
With the same setup, if the detector has false positives in the first 5 epochs and classifies the benign process correctly in the next 10 epochs, the effective slowdown is 26\%. To configure the level of slowdowns tolerable, \ourtool{} supports a user-specified limit on the minimum share of a resource available to a process, thereby limiting the slowdowns incurred. This configurability provides a trade-off between security and performance, as limiting overheads can allow a higher level of attack progress.

\section{Case Studies and Results}\label{sec:results}

For evaluation, we present four case studies, including various micro-architectural attacks~\cite{Aciimez:2010:instncache,Chakraborty:2022:TSA,Gras:2018:TLB,Maurice:2017:cjag,Osvik:2006:cacheFlushing,Yuval:2018:Mastik}, the rowhammer attack~\cite{Mutlu:2014:Rowhammer}, ransomware~\cite{bware_ransomware:2024,gonnacry_ransomware:2024,open_ransomware:2024,raasnet_ransomware:2024,randomware_ransomware:2024} and cryptominers~\cite{Papadopoulos:2018:cryptominers}. These case studies use different detectors based on existing works that have been augmented with~\ourtool{}.  For instance, the detectors for micro-architectural attacks use statistical models similar to~\cite{Payer:2016:HexPADS}, while the detector for ransomware uses time-series deep learning (DL) models similar to the ones used in~\cite{Alam:2017:HPCtoRescue,Chiapetta:2016:hpcdetection,Gulmezoglu:2019:FortuneTeller}. Table~\ref{tab:attacklist} describes the configurable aspects of~\ourtool{} for each attack along with the user specification, such as the progress metric calculation, penalty and compensation assessment functions ($\mathcal{F}_p \text{ and } \mathcal{F}_c$), and the actuator ($\mathcal{A}$). We present the details for each of these case studies in the next section. 

We evaluate~\ourtool{} on three platforms. 
First, an Intel Core i7-7700 processor. Second, an Intel i9-11900 processor, both running Ubuntu 20.04 on a Linux kernel version 4.19.265.
Third, an Intel Core i7-3770 processor with the Ivy Bridge micro-architecture and the Linux kernel version 4.19.2 with Ubuntu 16.04 operating system.

\begin{table*}[!t]
    \centering
    \caption{\small Case studies to evaluate~\ourtool{}. For each attack, we have the details of~\ourtool{} implementation, such as the progress metric, the detector ${\mathcal{D}}$ augmented by~\ourtool{}, penalty assessment function ($\mathcal{F}_p$), compensation assessment function ($\mathcal{F}_c$), and the actuator ($\mathcal{A}$).  }
    \begin{tabular}{|M{0.12\textwidth}|M{0.25\textwidth}|M{0.13\textwidth}|M{0.08\textwidth}|M{0.08\textwidth}|M{0.08\textwidth}|M{0.09\textwidth}|}
        \hline 
        \multirow{3}{*}{\textbf{Case Study}} & \multirow{3}{*}{\textbf{Attack(s)}} & \multicolumn{5}{c|}{\textbf{\ourtool{} implementation}}   \\
        \cline{3-7}
        &  & \textbf{Progress}  & \textbf{Detector augmented} & \textbf{$\mathcal{F}_p$} & \textbf{$\mathcal{F}_c$} & \textbf{Actuator ($\mathcal{A}$)} \\
        \hline
        \multirow{7}{*}{\shortstack{Micro-architectural\\ attacks}} & L1-D cache attack on AES~\cite{Osvik:2006:cacheFlushing} & Guessing entropy~\cite{massey:94:guessing} & \multirow{7}{*}{\shortstack{Statistical,\\ HPC-based}} & \multirow{7}{*}{\shortstack{Incremental \\(Equation~\ref{eqn:linear_penalty})}} & \multirow{7}{*}{\shortstack{Incremental\\ (Equation~\ref{eqn:linear_compensation})}} & \multirow{7}{*}{\shortstack{OS-Scheduler\\ based\\ (Equation~\ref{eqn:actuator})}} \\
             \cline{2-3}
        & L1-I cache attack on RSA \cite{Aciimez:2010:instncache} & Error rate & & & &\\
         \cline{2-3}
        & Load-Store Buffer covert channel~\cite{Chakraborty:2022:TSA}  & Error rate & & & &\\ 
         \cline{2-3}
         & CJAG high-speed covert channel \cite{Maurice:2017:cjag} & Bits transmitted & & & &\\
         \cline{2-3}
        & LLC covert channel \cite{Yuval:2018:Mastik} & Bits transmitted & & & &\\
         \cline{2-3}
        & TLB covert channel \cite{Gras:2018:TLB}  & Bits transmitted & & & &\\
        \hline      Rowhammer~\cite{Mutlu:2014:Rowhammer} & Rowhammer attack~\cite{rowhammer:2020:github} & Bits flipped & Statistical, HPC-based & Incremental & Incremental & {\shortstack{OS-Scheduler\\ based}}\\
        \hline
        Ransomware & Open-sourced samples~\cite{bware_ransomware:2024,gonnacry_ransomware:2024,open_ransomware:2024,raasnet_ransomware:2024,randomware_ransomware:2024} & Bytes encrypted & DL model, HPC-based & Incremental & Incremental & Cgroup based\\
        \hline 
        Cryptominer & Open-sourced samples~\cite{Papadopoulos:2018:cryptominers} & Hashes computed & Statistical, HPC-based & Incremental & Incremental & Cgroup based\\ 
        \hline 
    \end{tabular}
    \label{tab:attacklist}
    %\vspace{-0.5cm}
\end{table*}

% \vspace{-0.2cm}
\subsection{Case Study: Micro-architectural Attacks}\label{sec:results_micro_arch}

Micro-architectural attacks are a potent class of attacks that aim to break the isolation guarantees provided by the hardware.  A micro-architectural attack uses shared
hardware resources to leak information across these isolation boundaries. They have been used in a variety of applications, such as creating covert channels~\cite{Maurice:2017:cjag}, retrieving secret keys of  ciphers
~\cite{Bernstein:2005:CacheAttack}, reading Operating System data~\cite{Kocher:2018:Spectre,Lipp:2018:Meltdown}, breaking Address Space Layout Randomization~\cite{Barresi:2015:CAIN} and leaking secrets stored in Trusted Execution Environments like SGX~\cite{Brasser:2017:SGXattack}        
and Trustzone~\cite{Zhang:2018:Trustzone}. In a typical micro-architectural attack, the attacker runs a program called the spy that contends with a victim program for shared hardware resources such as a common cache memory%~\cite{ Aciimez:2010:instncache, Gruss:2015, Lipp:2016:ARMageddon, Liu:2015:LLC,percival:05}
~\cite{Aciimez:2010:instncache, Gruss:2015}, Branch Prediction Unit (BPU)%~\cite{Aciicmez:2007:BPU, Evtyushkin:2018:Branchscope}
~\cite{Evtyushkin:2018:Branchscope}, or Translation Lookaside Buffer (TLB)~\cite{Gras:2018:TLB}. The contention affects the spy's execution time in a manner that correlates with the victim's execution. If the victim's execution pattern happens to depend on secret data, then the correlation can be used to reveal it. Similarly, using the differences in execution time, two processes can establish a covert channel to transmit and receive bits~\cite{Chakraborty:2022:TSA,Gras:2018:TLB,Maurice:2017:cjag,Yarom:2014:flushReloadLLC}.

{\flushleft \bf Detector and assessment functions.} We augment a statistical detector using measurements from hardware performance counters (HPCs). Similar detectors have been presented in~\cite{Aweke:2016:anvil,Briongos:2018:cacheShield,Chiapetta:2016:hpcdetection,Mushtaq:2018:NightsWatch,Zhang:2016:cloudRadar} to classify malicious processes. To calculate the threat index, we use the following assessment functions. 

% \vspace{-0.8cm}
\begin{align}
    \mathcal{F}_P({\tt P}_{i-1}^{\tt t}) = {\tt P}_{i-1}^{\tt t} + 1\label{eqn:linear_penalty}\\
    \mathcal{F}_C({\tt C}_{i-1}^{\tt t}) = {\tt C}_{i-1}^{\tt t} + 1\label{eqn:linear_compensation}
\end{align}
% \vspace{-0.4cm}

The penalty function ensures that every time the detector classifies the process ${\tt t}$ as malicious, the penalty increases linearly, thereby increasingly throttling system resources. Similarly, the compensation function provides a mechanism for falsely classified benign programs to recover by increasing the available resources.

\begin{figure*}[!t]
\captionsetup[subfloat]{farskip=0.5pt,captionskip=0.5pt}
\centering
\begin{tabular}{ccc}

\subfloat[\small L1-data cache attack on AES]{\includegraphics[width=0.3\textwidth]{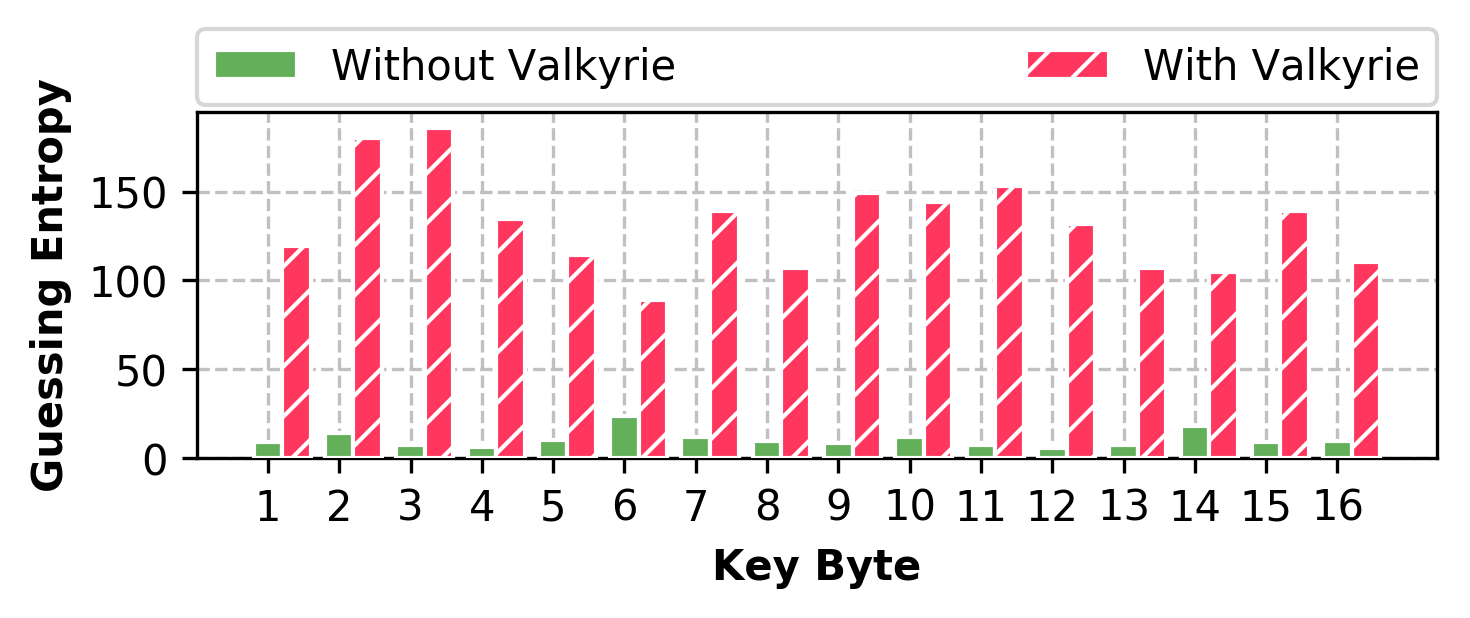}\label{fig:resAES}} &
\subfloat[\small L1-instruction cache attack on RSA]{\includegraphics[width=0.3\textwidth]{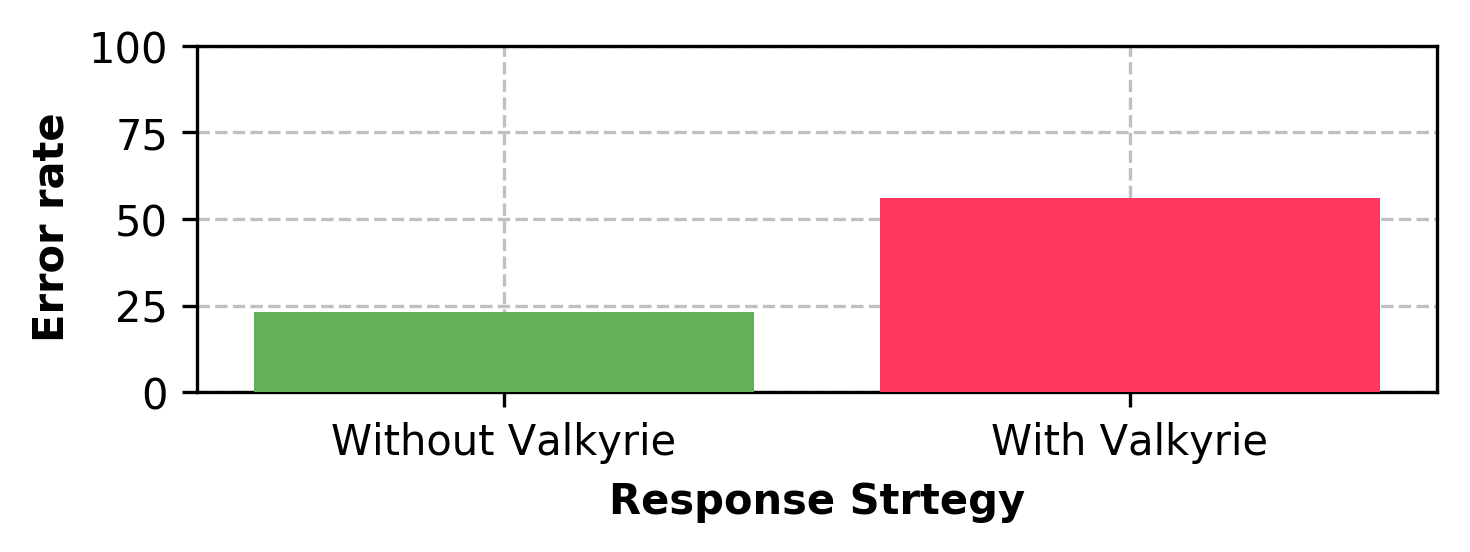}\label{fig:resL1instn}} &
\subfloat[\small TSA Covert Channel]{\includegraphics[width=0.3\textwidth]{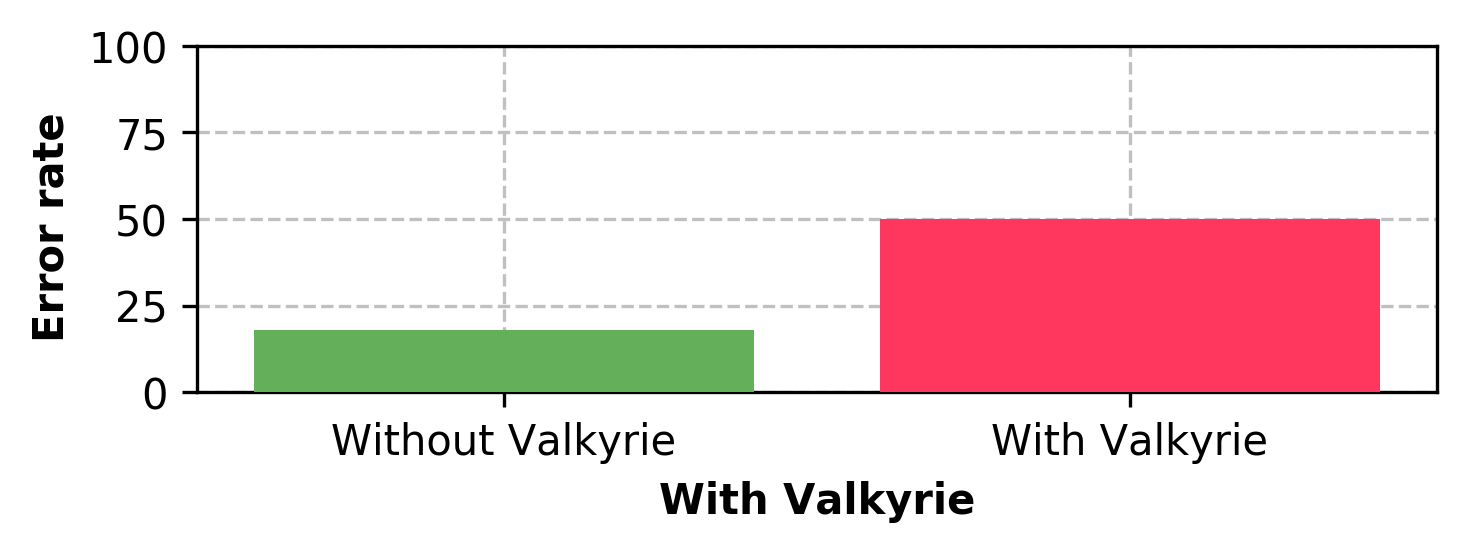}\label{fig:TSA}}\\

\subfloat[\small CJAG cache covert channel]{\includegraphics[width=0.3\textwidth]{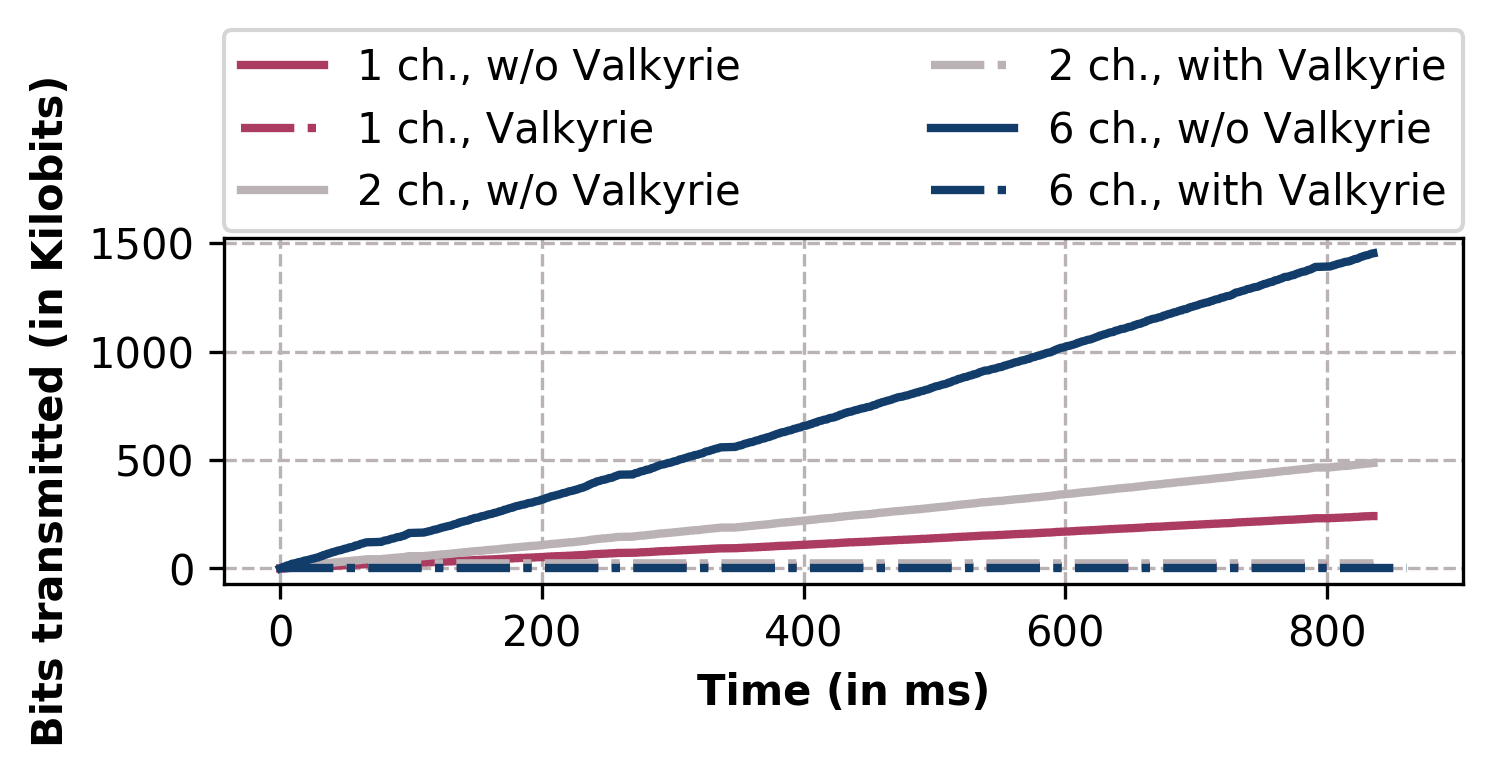}\label{fig:resCJAG}} & 

\subfloat[\small LLC covert channel]{\includegraphics[width=0.3\textwidth]{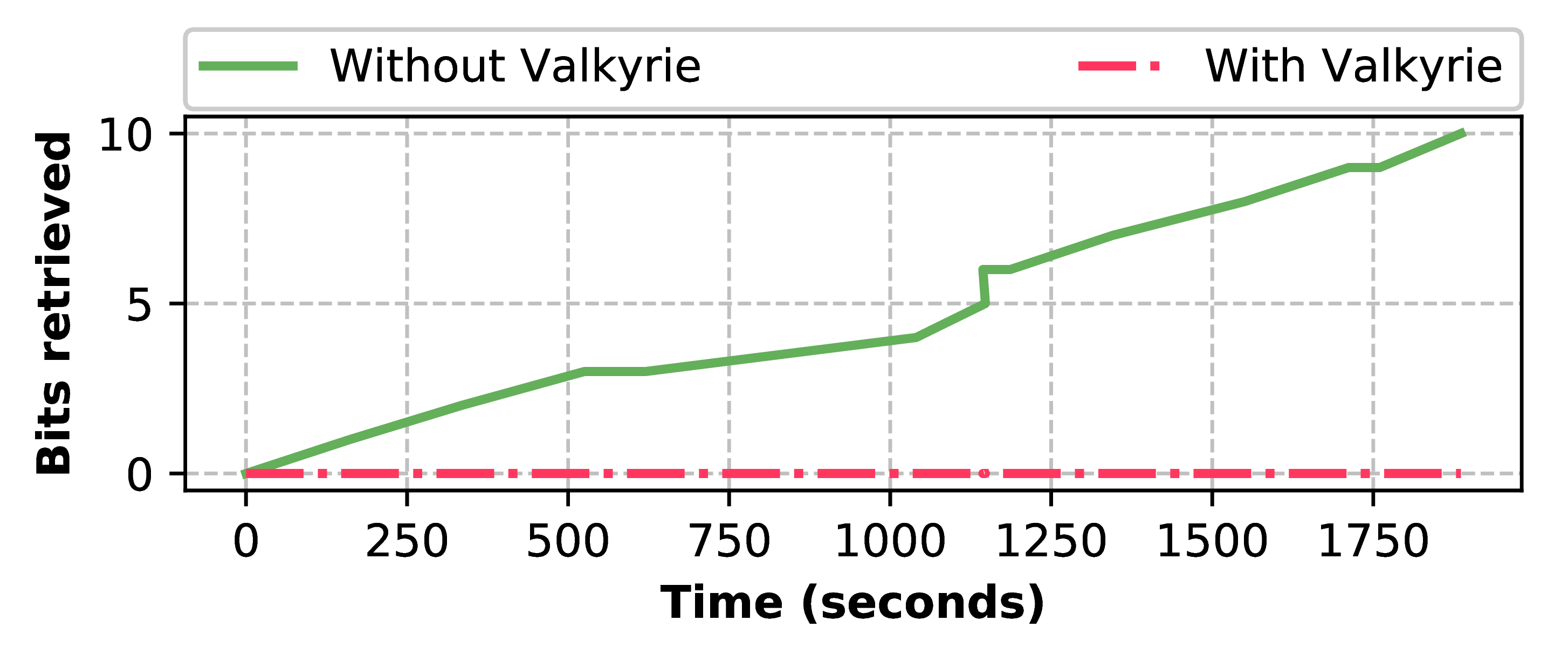}\label{fig:resLLC}} &

\subfloat[\small TLB covert channel]{\includegraphics[width=0.3\textwidth]{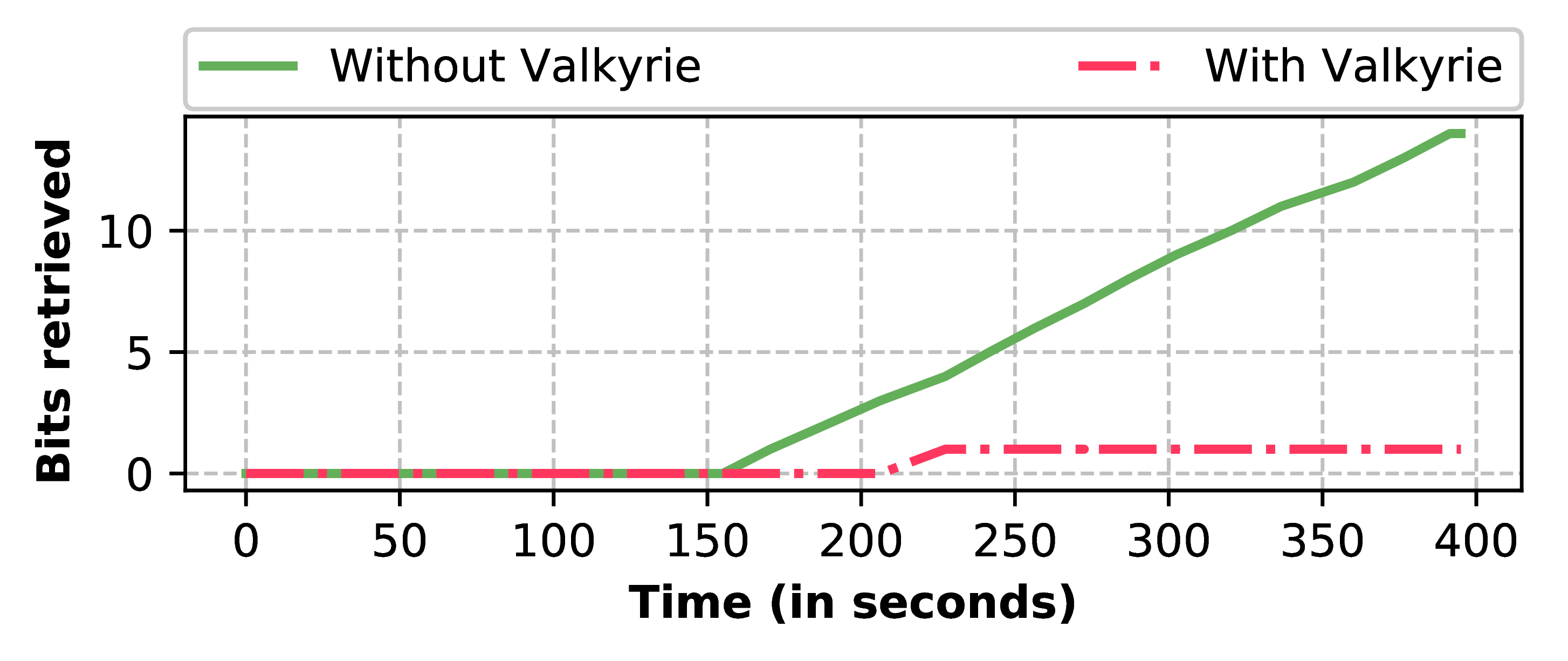}\label{fig:resTLB}} \\

\end{tabular}
% \vspace{-0.2cm}
\caption{\small The impact of~\ourtool{} on the progress of various micro-architectural attacks. }\label{fig:attack_eval}
% \vspace{-0.5cm}
\end{figure*}

{\flushleft \bf OS scheduler-based actuator function.} A common characteristic of micro-architectural attacks is the dependence on the available CPU time. We leverage this by using an actuator function that controls the CPU time available to a process. The actuator function modifies the OS scheduler such that the execution time of processes is dependent on their threat index values. The Linux kernel, since Version 2.6., incorporates a Completely Fair Scheduler (CFS), which tries to achieve the ideal multitasking environment where processes with equal priorities receive the same share of CPU time for execution, called {\em timeslice}. The {timeslice} allocated to a process ${\tt t}$, denoted $\Delta_{\tt ts}^{\tt t}$, is a fraction of a  predefined value called {\em targeted latency} ($\Delta_{\tt tl}$). When multiple processes compete for CPU time, the scheduler  allocates timeslices in proportion to a metric called {\em weight} of the process as follows

% \vspace{-0.2cm}
\begin{equation}
    \Delta_{\tt ts}^{\tt t} = \Delta_{\tt tl} \times  \frac{{\tt w}_{i}^{\tt t}}{\sum_{\tt processes} {\tt w}_{i}} = \Delta_{\tt tl} \times {\tt s}_{i}^{\tt t} \enspace,
    \label{eqn:ts}
\end{equation}

where ${\tt w}_{i}^{\tt t}$ is the weight of the process ${\tt t}$, ${\sum_{\tt processes} {\tt w}_i}$  is the sum of weights of all the processes sharing the CPU, and ${\tt s}_{i}^{\tt t}$ is the {\em relative weight} of process ${\tt t}$. When a process starts execution, its {weight} takes a default value, which lies in the middle of 40 discrete levels. The difference in weights at two consecutive levels $\gamma$, $(0 < \gamma < 1)$ is determined by the OS scheduler at design time. A higher weight value for a thread implies a larger timeslice and a higher frequency of getting scheduled for execution, and hence more CPU time. The actuator function $\mathcal{A}$ maps the weight level of the process to its threat index, given by,

% \vspace{-0.5cm}
\begin{align}
    {\tt s}_{i}^{\tt t} &= \mathcal{A}({\tt s_{i-1}^{\tt t}}, \Delta {\tt T}_{i, 1}^{\tt t}) \nonumber\\ 
    &= \begin{cases} 
    {\tt s}_{i-1}^{\tt t} - \gamma \times ({\tt s}_{i-1}^{\tt t})\times \Delta {\tt T}_{i, 1}^{\tt t},\quad \Delta {\tt T}_{i, 1}^{\tt t} > 0\\
    {\tt s}_{i-1}^{\tt t} + \gamma \times ({\tt s}_{i-1}^{\tt t})\times \Delta {\tt T}_{i, 1}^{\tt t},\quad \Delta {\tt T}_{i, 1}^{\tt t} \leq 0 \enspace,
    \end{cases}\label{eqn:actuator}
\end{align}

where ${\gamma}$ determines the amount of fall in the weight with every increase in the threat index, ${\tt s}_{i-1}^{\tt t}$ is the relative weight of process ${\tt t}$ in the $i$-th epoch, and $\Delta {\tt T}_{i,1}^{\tt t} =  {\tt T}_{i}^{\tt t} - {\tt T}_{i-1}^{\tt t}$. In our evaluation platforms, $\gamma = 0.1$, which means that every rise in threat index decreases the relative weight of the process by 10\% until it reaches the minimum value ${\tt s_{MIN}}$. Similarly, when a process is in the suspicious state, every drop in the threat index increases the process's relative weight by 10\% until it goes back to the normal state (Equation~\ref{eqn:actuator}).

\begin{figure*}[!t]
\captionsetup[subfloat]{farskip=0.5pt,captionskip=0.25pt}
\centering
\begin{tabular}{M{0.65\textwidth}M{0.3\textwidth}}

\subfloat[\small]{\includegraphics[width=0.68\textwidth]{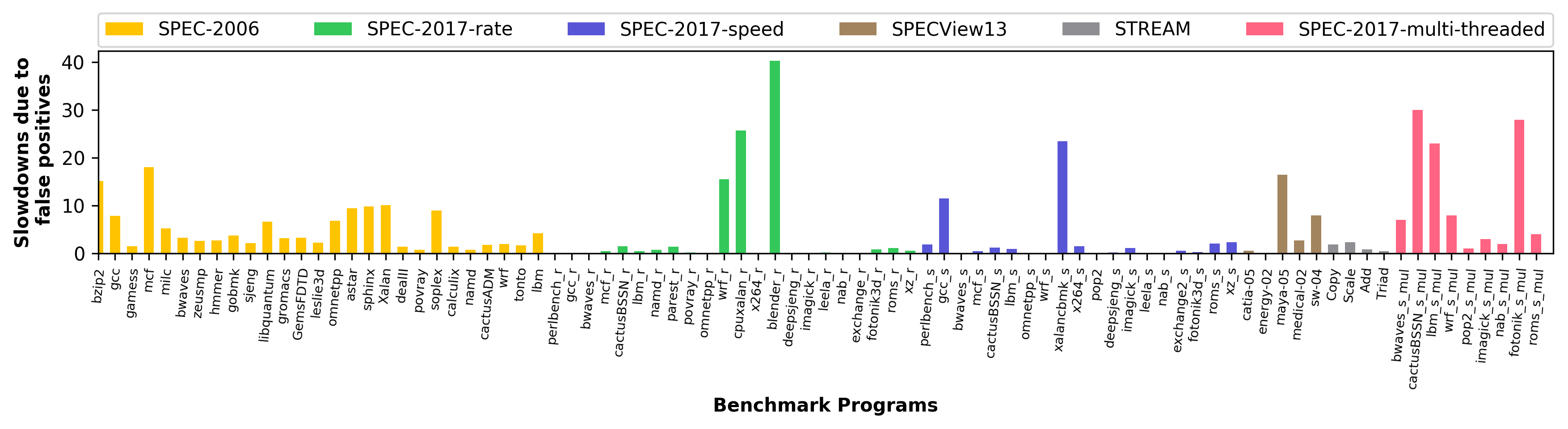}\label{fig:perfOverhead}} &
\subfloat[\small]{\includegraphics[width=0.3\textwidth]{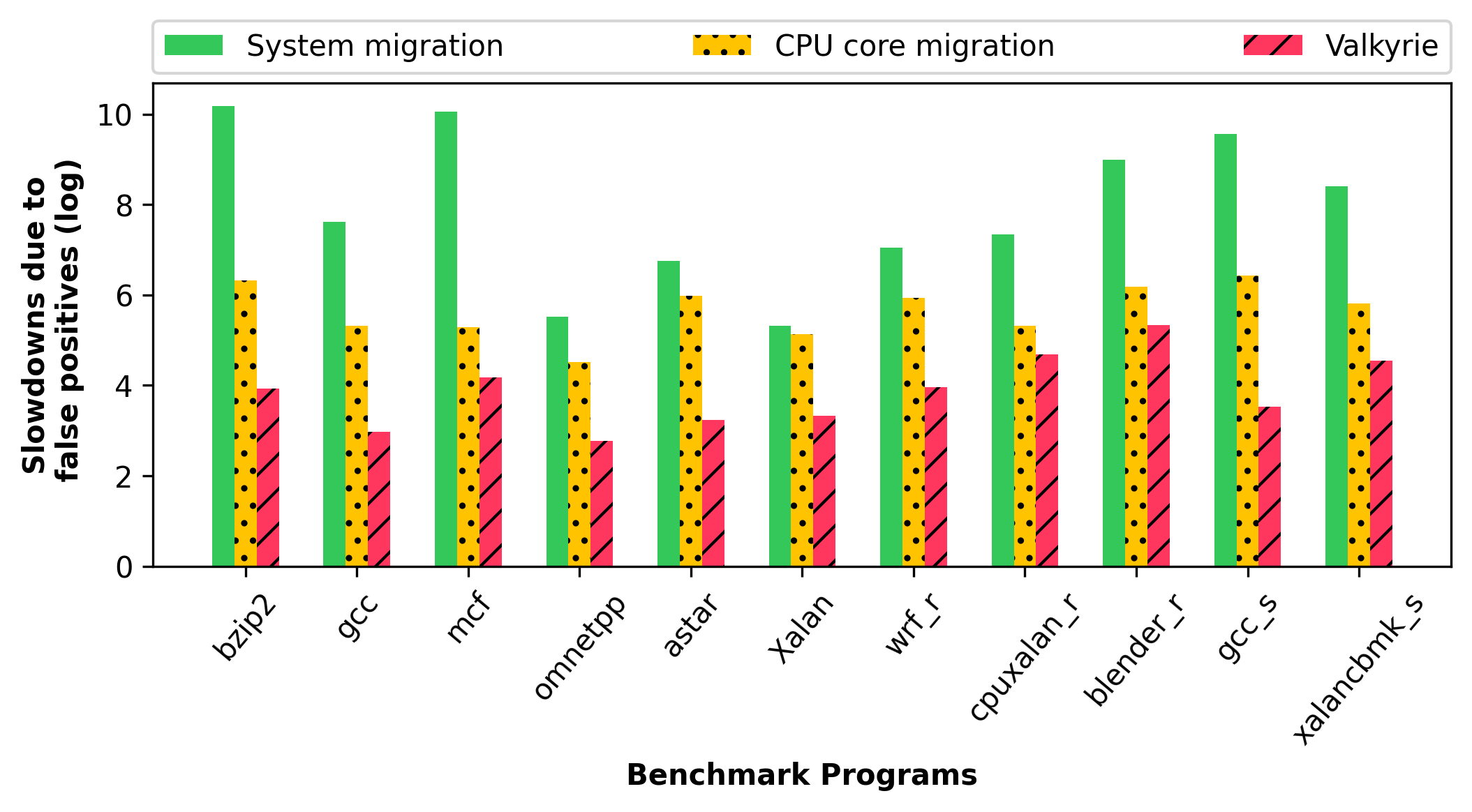}\label{fig:migration_vs_valkyrie}}\\

\end{tabular}
% \vspace{-0.4cm}
\caption{\small (a) Slowdowns with~\ourtool{} on programs from different benchmark suites including SPEC-2006~\cite{SPEC2006}, SPEC-2017~\cite{spec2017}, SPECView13~\cite{specviewperf13}, STREAM~\cite{McCalpin:2007:stream} and multi-threaded SPEC-2017~\cite{spec2017} due to false positives. (b) Slowdowns due to false positives with different post-detection strategies for micro-architectural attacks, i.e., system migration, CPU core migration, and~\ourtool{}.}
% \vspace{-0.6cm}
\end{figure*}

{\flushleft \bf Throttling micro-architectural attacks with~\ourtool{}.} Our evaluation covers various micro-architectural attacks that target different micro-architectural components as listed in Table~\ref{tab:attacklist}. These attacks include an L1 data cache attack on AES~\cite{Osvik:2006:cacheFlushing}, an L1 instruction cache attack on RSA~\cite{Aciimez:2010:instncache}, a cache-agnostic covert channel using the Load and Store buffers~\cite{Chakraborty:2022:TSA}, a high-speed covert channel called CJAG~\cite{Maurice:2017:cjag}, a LLC covert channel~\cite{Yuval:2018:Mastik}, and a TLB covert channel~\cite{Gras:2018:TLB}.

Fig.~\ref{fig:attack_eval} describes the impact of~\ourtool{} on the progress of these attacks after they have been detected and transitioned to the suspicious state. To understand the effectiveness of~\ourtool{},
we use different metrics to represent the attack's progress $\mathrm{B}_i^{\tt t}(R_i^{\tt t})$ (as shown in Table~\ref{tab:attacklist}). For example, to quantify the progress of the L1-D cache attack that performs key recovery on a T-table implementation of AES, we use the Guessing Entropy~\cite{massey:94:guessing}. The Guessing Entropy metric defines the number of possible values for the key byte. A Guessing Entropy of 128 indicates that the attacker has no significant benefits from the timing measurements, as compared to a random guess. As the attack progresses and performs more timing measurements, the Guessing Entropy decreases. 
As shown in Fig.~\ref{fig:resAES},~\ourtool{} increases the guessing entropy of the attack from 10 to 131, thereby thwarting the attack. For the L1-instruction cache attack on RSA and the TSA load-store buffer covert channel, we quantify the progress based on the error in guessing 1-bit of the key correctly. With~\ourtool{}, the error rate for both these attacks increases to more than 50\% (Fig.~\ref{fig:resL1instn} and Fig.~\ref{fig:TSA}), rendering the attacks on par with randomly guessing the key bits.

For the covert channels using the LLC~\cite{Yuval:2018:Mastik,Maurice:2017:cjag} and TLB~\cite{Gras:2018:TLB}, we represent the progress by the number of bits transmitted.  The Cache-based Jamming Agreement (CJAG) is the fastest micro-architectural covert channel~\cite{Maurice:2017:cjag} to date. CJAG supports multiple communication channels in the LLC, noise characterization, and error correction to retrieve bits. During initialization, the sender and receiver identify cache sets that serve as channels for communication. Post initialization, a 2-way communication protocol is used to transmit bits from the sender to the receiver with speeds of over 40KB/second. Fig.~\ref{fig:resCJAG} describes the impact of~\ourtool{} on CJAG with different configurations of communication channels. After the channels are throttled, no bits are transmitted, clamping the information leaked from sender to receiver. We observe that as the number of channels increases, the bits transferred by CJAG  decrease (Fig.~\ref{fig:resCJAG}). This is because a large number of channels would require a longer initialization period, giving~\ourtool{} time to throttle the channel before bits are transmitted.  Similarly, the covert channels using LLC~\cite{Yuval:2018:Mastik} and TLB~\cite{Gras:2018:TLB} see a drastic fall in the number of bits communicated after getting throttled by~\ourtool{} (Fig.~\ref{fig:resLLC} and~\ref{fig:resTLB}).

\begin{figure*}[!t]
\captionsetup[subfloat]{farskip=0.5pt,captionskip=0.5pt}
\centering
\begin{tabular}{ccc}

\subfloat[\small]{\includegraphics[width=0.31\textwidth]{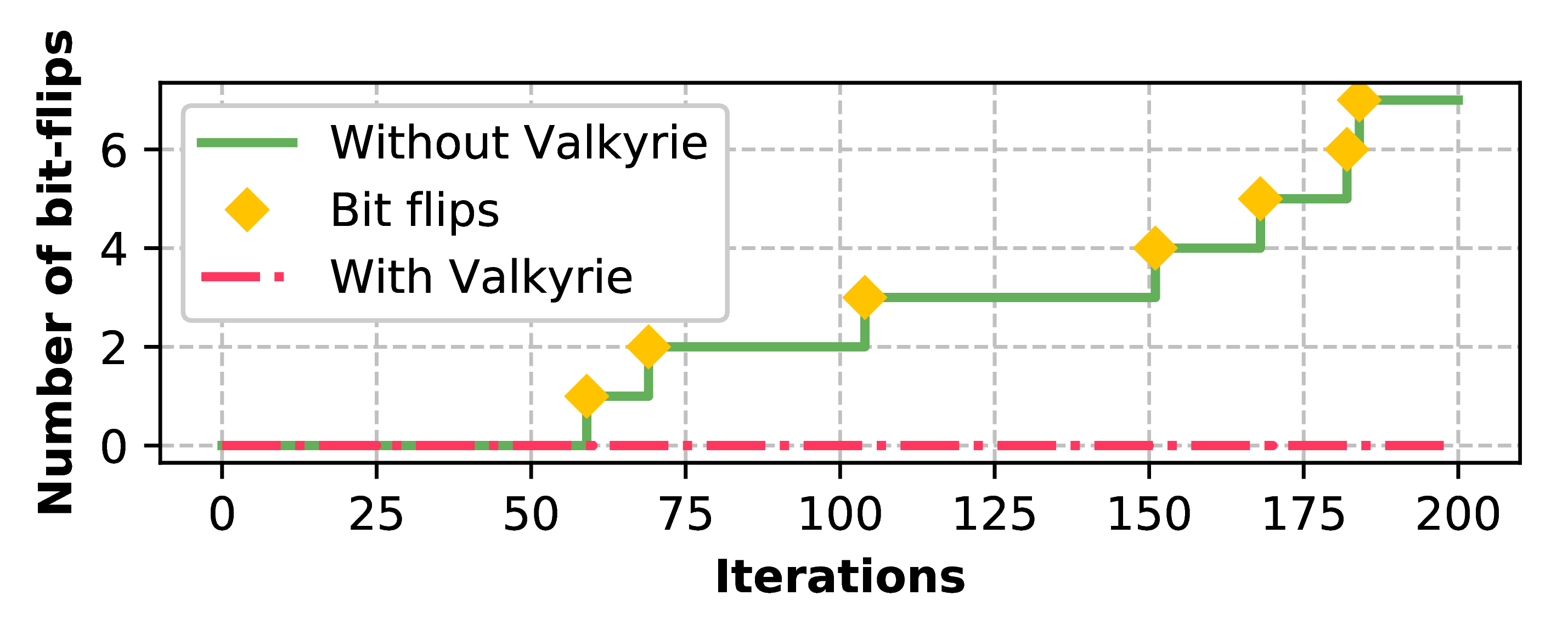}\label{fig:resRowhammer}} &
\subfloat[\small]{\includegraphics[width=0.3\textwidth]{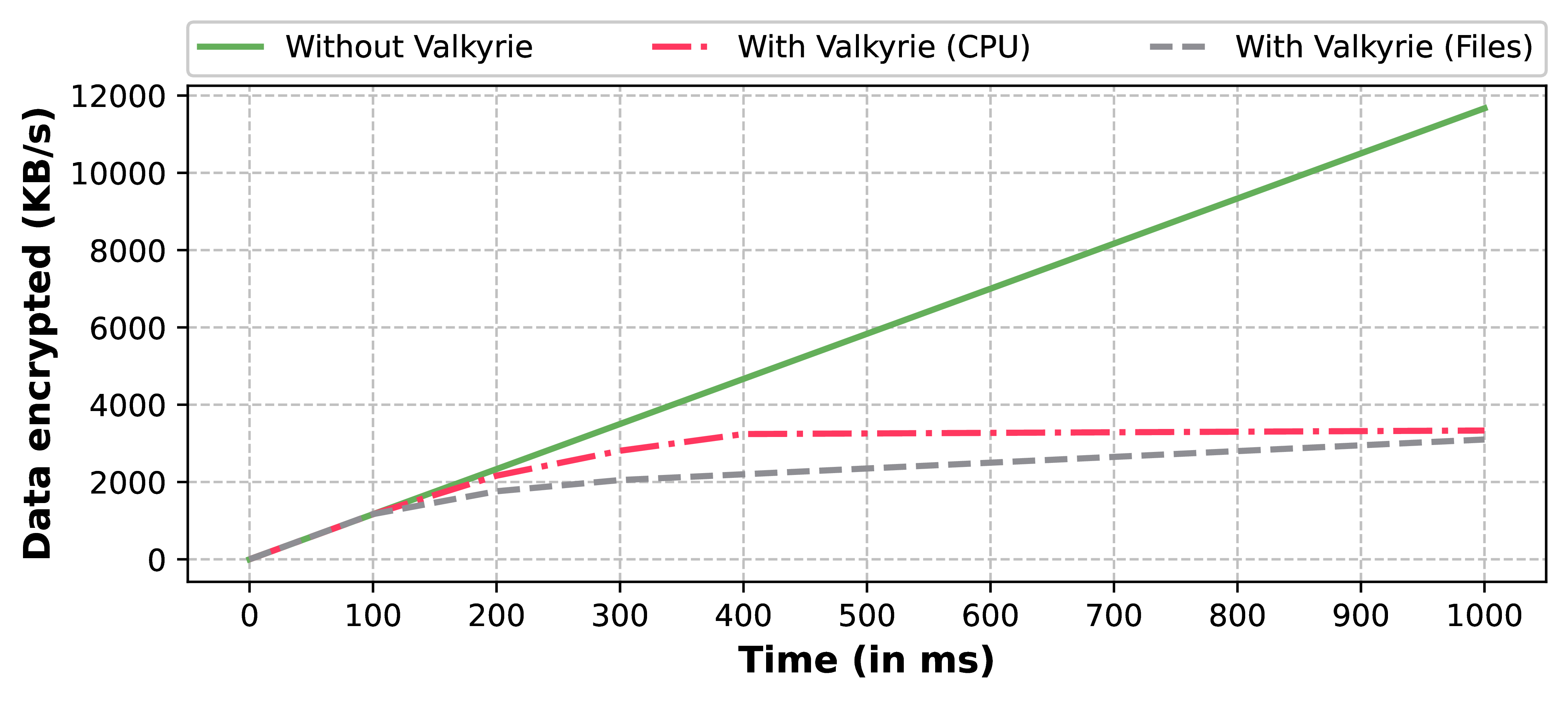}\label{fig:ransomware_results}} &
\subfloat[\small]{\includegraphics[width=0.31\textwidth]{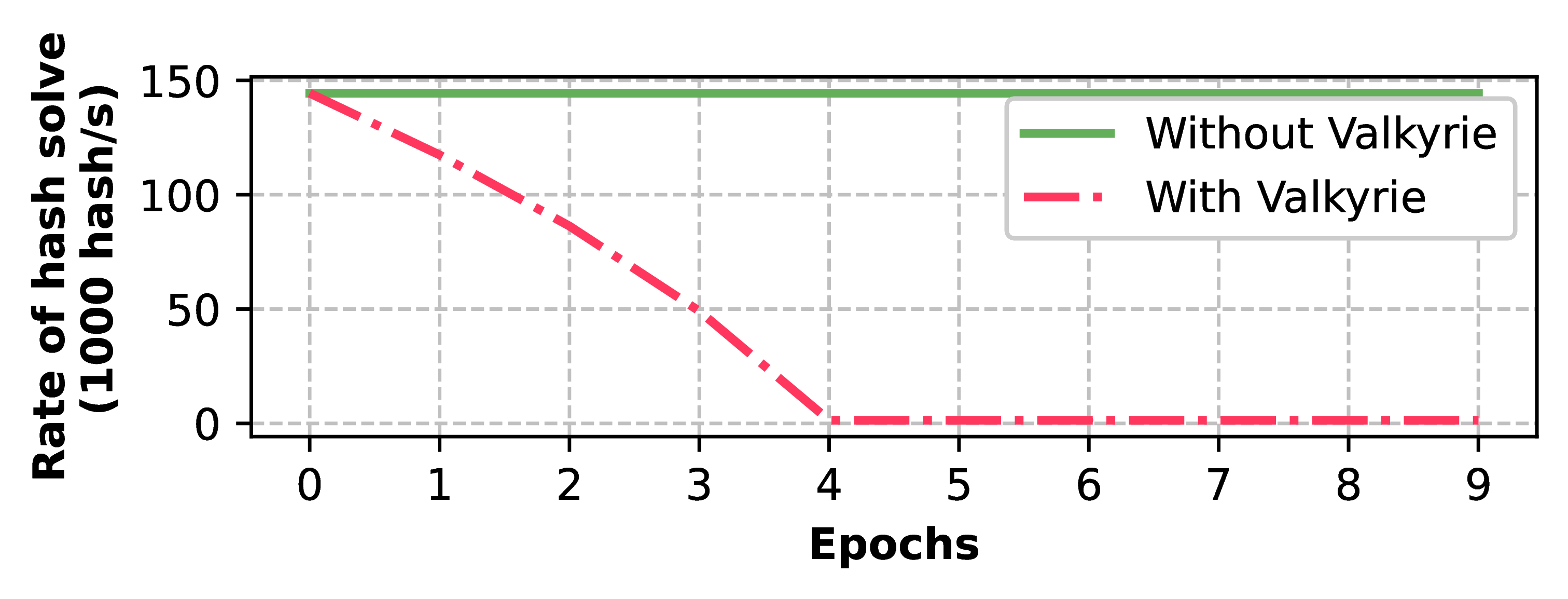}\label{fig:cryptominers_results}}\\

\end{tabular}
% \vspace{-0.2cm}
\caption{\small (a) The impact of~\ourtool{} on the number of bits flipped by the rowhammer~\cite{rowhammer:2020:github} attack. By throttling the CPU time available to the attack,~\ourtool{} induces a 100\% slowdown evaluated in a day of attack execution. (b) The average rate of encryption of data with and without~\ourtool{}. (c) The average rate of hash computations by cryptominers with and without~\ourtool{}.}
% \vspace{-0.7cm}
\end{figure*}

% {\vspace{-0.2cm}}
{\flushleft \bf Slowdowns due to false positives.} As discussed in Section~\ref{sec:slowdowns},~\ourtool{} can induce slowdowns in benign processes by throttling resources when the detector has false positives. We evaluate these slowdowns with multiple benchmark suites namely SPEC-2006~\cite{SPEC2006}, SPEC-2017~\cite{spec2017}, SPECViewperf-13~\cite{specviewperf13}, STREAM~\cite{McCalpin:2007:stream} and the multi-threaded SPEC-2017~\cite{spec2017} benchmarks. SPEC-2006 and SPEC-2017 are CPU benchmark suites with different integer and floating-point programs like Machine Learning algorithms. SPECViewperf-13 is a collection of graphics-oriented benchmark programs, while STREAM is designed to perform memory-intensive tasks. The multi-threaded SPEC-2017 suite has floating-point multi-threaded programs that spawn 4 threads during the evaluation.

A simple statistical detector effectively demonstrates the capabilities of~\ourtool{}, as it is expected to have a higher frequency of false positives compared to more complex detectors. For instance, the detector used for micro-architectural attack detection classifies programs from the SPEC-2006 suite as malicious in 4\% of the epochs, on average. Fig.~\ref{fig:perfOverhead} presents the slowdowns due to~\ourtool{} incurred by these benchmark programs. Out of the 77 single-threaded programs evaluated, 60 have slowdowns of less than {5\%}, while 35 have less than 1\% slowdowns. The overall average across all benchmarks is 1\% (geometric mean) or 2.8\%(arithmetic mean) for single-threaded programs, while the maximum slowdown incurred is 40.3\%. We further evaluate the benchmarks on two other platforms, namely, Intel i7-7700 and Intel i9-11900, which have an average runtime slowdown of 2.2\% and under 1\%, respectively, as shown in Table~\ref{tab:processor_overheads}.  On the other hand, multi-threaded programs incur an average slowdown of about 6.7\%.

\begin{table}[!t]
    \centering
    % \vspace{0.2cm}
        \caption{\small Average (geometric mean) slowdowns with~\ourtool{} on SPEC-2017~\cite{spec2017} programs due to false positives on different execution environments.}
     \begin{tabular}{|M{0.15\columnwidth}|M{0.5\columnwidth}|M{0.17\columnwidth}|}
        \hline
        {\bf Processor} & \textbf{OS and Linux Kernel} & \textbf{Slowdowns}\\
        \hline
        i7-3770 & Ubuntu 16.04, Linux 4.19.2 & 1\%\\
        \hline
        i7-7700 & Ubuntu 20.04, Linux 4.19.265 & 2.2\%\\
        \hline
        i9-11900 & Ubuntu 20.04, Linux 4.19.265 & $<$1\%\\
        \hline
    \end{tabular}
    \label{tab:processor_overheads}
\end{table}

In contrast to contemporary post-detection responses
~\cite{Alam:2017:HPCtoRescue,Briongos:2018:cacheShield,Chiapetta:2016:hpcdetection,Kulah:2019:SpyDetector,Mushtaq:2018:NightsWatch,Mushtaq:2020:whisper,Mushtaq:2021:transit_guard,Nomani:2015:schedHPC,Payer:2016:HexPADS,Zhang:2016:cloudRadar}, all falsely classified benign processes recover and are not adversely affected. As an example, let us consider {\tt blender\_r},  a 3D rendering program, which is falsely classified by the detector in 30\% of the epochs. Assuming the same detector and a termination based response~\cite{Alam:2017:HPCtoRescue,Briongos:2018:cacheShield,Chiapetta:2016:hpcdetection,Kulah:2019:SpyDetector,Mushtaq:2018:NightsWatch,Mushtaq:2020:whisper,Mushtaq:2021:transit_guard},
{\tt blender\_r} would have been terminated with a probability of 0.3. Improving the detection algorithm cannot completely prevent the termination. In contrast to this,~\ourtool{} throttles the program, resulting in a slowdown of 25\% (the highest slowdown observed across all single-threaded benchmarks). Another response strategy for micro-architectural attack detection is to migrate the detected processes to a different  CPU core or a different system via the network~\cite{Nomani:2015:schedHPC,Zhang:2016:cloudRadar}. With migration schemes, the slowdowns for {\tt blender\_r} would have as high as 10X as that of~\ourtool{}. Fig.~\ref{fig:migration_vs_valkyrie} compares the slowdowns of benchmark programs with different migration techniques.
We observe that on each detection, the migration of a process to a different CPU core in the same machine has 1.5X more overheads, while the migration of the process to a different system performs 4X slower than the response from~\ourtool{}, on average. Thus,~\ourtool{} provides a mechanism for a reactive post-detection response, even with a highly simplistic detector.

\subsection{Case Study: Rowhammer Attack}\label{sec:results_rowhammer}

The rowhammer attack~\cite{Mutlu:2014:Rowhammer} flips the bits stored in a DRAM cell by frequently accessing the adjacent cells in a loop. To this end, the attacker iteratively performs memory accesses to the DRAM while flushing the cache to ensure that each load is fetched from the memory. These bit flips induced by rowhammer have been used for various exploits such as gaining kernel privileges~\cite{Seaborn:2016:rowhammer_implementation}, breaking isolation between VMs~\cite{Razavi:2016:RowHammerVM}, and 
compromising cryptographic implementations~\cite{Razavi:2016:RowHammerVM}.

We evaluate~\ourtool{} by augmenting it to an HPC-based statistical detector, similar to~\cite{Aweke:2016:anvil}. We use a linear penalty and compensation function. The rowhammer attack uses both the CPU time and the memory resources. However, in each iteration, the attack only accesses a small number of addresses. Thus, we throttle the execution using the actuator function shown in Equation~\ref{eqn:actuator}. In our experiments, we use a popular open-sourced implementation of the attack~\cite{rowhammer:2020:github}. On average, this attack induces a bit flip in every 29 iterations on our evaluation DRAM chip, Transcend  DDR3-1333 645927-0350. Fig.~\ref{fig:resRowhammer} demonstrates the throttling of the rowhammer attack in the suspicious state with~\ourtool{} such that no bit-flips are observed even after a day of execution. Thus, in our evaluation, the attack incurs a 100\% slowdown with~\ourtool{} before termination.

% \vspace{-0.15cm}
\subsection{Case Study: Ransomware Attacks}\label{sec:results_ransomware}
% \vspace{-0.1cm}

Ransomware attacks are a class of malware that encrypts the filesystem of infected electronic devices such as consumer devices or enterprise systems~\cite{sonicwall_threat_report2024,thales_threat_report2024}, rendering them useless. A popular example is the Wannacry ransomware attack~\cite{wannacry_blog:2017}, which affected 400K devices in over 150 countries. The easy availability of attack programs to malevolent actors via businesses providing Ransomware-as-a-Service (RaaS)~\cite{Meland:2020:RaaS} has exacerbated the spread of these attacks in recent years. 

We augment a detector based on deep learning approaches (similar to~\cite{Alam:2017:HPCtoRescue,Mushtaq:2018:NightsWatch,Mushtaq:2021:transit_guard}) with~\ourtool{}. We use a Long Short-Term Memory (LSTM) model trained on the time-series HPC measurements from a dataset of 67 ransomware from open-sourced repositories~\cite{bware_ransomware:2024,gonnacry_ransomware:2024,open_ransomware:2024,raasnet_ransomware:2024,randomware_ransomware:2024} and benign programs from the SPEC-2006~\cite{SPEC2006} benchmark suite. The LSTM neural network has an input layer of 20 nodes, a hidden layer of 8 nodes, and an output layer with a sigmoid activation function.  We use a linear penalty and compensation assessment function (Equation~\ref{eqn:linear_penalty} and~\ref{eqn:linear_compensation}). 

We quantify the progress of ransomware attacks with the amount of data encrypted. Since ransomware attacks utilize the CPU for encrypting the filesystem, we demonstrate actuator functions that throttle both these resources in the suspicious state. 
Fig.~\ref{fig:ransomware_results} shows the data encrypted by the evaluated ransomware attacks on average with and without~\ourtool{}. In our experiment, each new measurement and inference takes 100ms. Without~\ourtool{}, these attacks can encrypt data with a rate of 11.67MB/second. With an actuator that throttles the available CPU time, the rate of progress drops to 152KB/second after the fifth epoch. For the filesystem, we use an actuator that halves the rate of file accesses every time there is an increase in the threat index. Thus, the attack's file access rate goes down from 7 files per epoch to 1 file per epoch. This brings down the rate of encryption to 1.5MB/second. The attack can be terminated at different points based on the user-specified detection efficacy. For instance, to achieve an F1-Score of 
0.85, our ANN takes 20 epochs. During this period,~\ourtool{} throttles the ransomware to restrict the encrypted data to about 3.5 MB as compared to 233 MB without~\ourtool{}.

\subsection{Case Study: Cryptominers}\label{sec:results_cryptominer}

Cryptominer attacks attempt to use the CPU resources of a victim's system with a financial motivation to \textit{mine} cryptocurrency. Typically, mining involves guessing a hash input that results in an output of a specific pattern. With the growing popularity of cryptocurrency, such attacks are growing rapidly~\cite{Papadopoulos:2018:cryptominers}.

We use an HPC-based statistical detector for detecting cryptominers similar to~\cite{Karapoola:2024:Sundew}. The penalty and compensation assessment functions are linear. Since cryptominers are computationally expensive, the actuator throttles the available CPU time upon detection. The average slowdown of cryptominers with~\ourtool{} is 99.04\% (Fig.~\ref{fig:cryptominers_results}) in the suspicious state. 

% \vspace{-0.2cm}
\section{Discussion}\label{sec:caveats}
% \vspace{-0.1cm}

In this section, we present a questionnaire outlining the scope of the paper and caveats associated with~\ourtool{}.

{\flushleft \bf Can~\ourtool{} improve the detection efficacy of detectors?} No. \ourtool{} is not an attack detector, and it can not directly influence the capabilities of a detector. Rather, it is a post-detection framework that can augment runtime detectors. The goal of~\ourtool{} is to reduce the adverse impacts of false positives while thwarting time-progressive attacks.

{\flushleft \bf Can~\ourtool{} counter adversarial attacks on detectors?} No. Adversarial attacks can evade detection by exploiting the limitations of the detection model. Such attacks have been shown to be effective against different statistical and machine learning approaches~\cite{Chakraborty:2018:adversarial_survey}. These attacks are a limitation of the underlying detector. Since~\ourtool{} enables responses only after detection, the susceptibility of a detector to adversarial attacks is out of scope for this paper. A possible solution is to use multi-level detection approaches as presented in~\cite{Ozsoy:2015:two_level_detections} before augmenting them with~\ourtool{}.

{\flushleft \bf Is~\ourtool{} limited to detectors using Hardware Performance Counters?} No. As shown in Fig.~\ref{fig:valkyrie_cycle}, in every epoch~\ourtool{} takes the inference from the detector for threat assessment and managing available system resources, agnostic to the low-level details of the detector. The case studies presented in Section~\ref{sec:results} use detectors based on existing works~\cite{Alam:2017:HPCtoRescue,Briongos:2018:cacheShield,Chiapetta:2016:hpcdetection,Gulmezoglu:2019:FortuneTeller,Karapoola:2024:Sundew,Mushtaq:2018:NightsWatch,Mushtaq:2020:whisper,Mushtaq:2021:transit_guard,Nomani:2015:schedHPC,Zhang:2016:cloudRadar,Payer:2016:HexPADS}, which typically make use of HPCs.

% \vspace{-0.1cm}
\section{Conclusion}\label{sec:conclusions}
% \vspace{-0.1cm}

A major shortcoming of real-time cyberattack detection is the detrimental impact of false positives. Existing research 
aims to reduce false positives by deploying complex detection algorithms, yet none can completely eliminate them, leading to lower productivity and usability of computer systems. 
~\ourtool{} augments detectors to mitigate the adverse impacts of false positives by throttling system resources until the detectors have sufficient confidence to terminate the program. 

By shifting focus from the detection algorithm to the response, \ourtool{} enables the use of lightweight detectors. This is particularly helpful for resource-constrained devices on which complex detection algorithms are impractical. Additionally,~\ourtool{} also enables users to configure security based on the application requirements, enhancing adoption across various domains. The paper opens up a new avenue for research dealing with the post-detection impacts of countermeasures and their applications.

\section*{Acknowledgement}
Our research work was partially funded by the European Union under Horizon Europe Programme – Grant Agreement 101070537 –
CrossCon, the European Research Council under the ERC
Programme - Grant 101055025 - HYDRANOS, the Deutsche Forschungsgemeinschaft (DFG) – SFB 1119 – 236615297, and the Information Security Education and Awareness (ISEA) initiative, Ministry of Electronics and Information Technology (MEiTY), Government of India. Any opinions, findings, conclusions, or recommendations expressed herein are those of the authors and do not necessarily reflect those of the European Union, the European Research Council, the Deutsche Forschungsgemeinschaft, or the Government of India.

\bibliographystyle{plain}
\bibliography{conference_101719}

\end{document}